\documentclass[fleqn,usenatbib]{mnras}

\usepackage{newtxtext,newtxmath}

\usepackage[T1]{fontenc}

\DeclareRobustCommand{\VAN}[3]{#2}
\let\VANthebibliography\thebibliography
\def\thebibliography{\DeclareRobustCommand{\VAN}[3]{##3}\VANthebibliography}

\usepackage{graphicx}	
\usepackage{amsmath}	

\graphicspath{{./}{figures/}}

\title[ATM: Systematic detection of outflows in LMXBs]{Ask The Machine: Systematic detection of wind-type outflows in low-mass X-ray binaries}

\author[D. Mata S\'anchez et al.]{D. Mata S\'anchez$^{1,2}$\thanks{E-mail: matasanchez.astronomy@gmail.com}, T. Mu\~noz-Darias$^{1,2}$, J. Casares$^{1,2}$, M. Huertas-Company$^{1,2,3}$, G. Panizo-Espinar$^{1,2}$  \\
$^{1}$Instituto de Astrof\'{i}sica de Canarias, E-38205 La Laguna, Tenerife, Spain \\
$^{2}$Departamento de astrof\'{i}sica, Universidad de La Laguna, E-38206 La Laguna, Tenerife, Spain \\
$^{3}$LERMA, Observatoire de Paris, CNRS, PSL, Universit\'e Paris-Cit\'e, France
}

\date{Accepted 2023 June 20. Received 2023 June 19; in original form 2023 May 18}

\pubyear{2023}

\begin{document}
\label{firstpage}
\pagerange{\pageref{firstpage}--\pageref{lastpage}}
\maketitle

\begin{abstract}
The systematic discovery of outflows in the optical spectra of low-mass X-ray binaries opened a new avenue for the study of the outburst evolution in these extreme systems. However, the efficient detection of such features in a continuously growing database requires the development of new analysis techniques with a particular focus on scalability, adaptability, and automatization. In this pilot study, we explore the use of machine learning algorithms to perform the identification of outflows in spectral line profiles observed in the optical range. We train and test the classifier on a simulated database, constructed through a combination of disc emission line profiles and outflow signatures, emulating typical observations of low-mass X-ray binaries. The final, trained classifier is applied to two sets of spectra taken during two bright outbursts that were particularly well covered, those of V404 Cyg (2015) and MAXI J1820+070 (2018). The resulting classification gained by this novel approach is overall consistent with that obtained through traditional techniques, while it simultaneously provides a number of key advantages over the latter, including the access to low velocity outflows. This study sets the foundations for future studies on large samples of spectra from low-mass X-ray binaries and other compact binaries.

\end{abstract}




\section{Introduction}

The field of X-ray binaries has grown steadily in the last 60 years, since the discovery of Sco X-1, the first and brightest stellar X-ray source in the night sky \citep{Giacconi1962}. Over 200 low-mass X-ray binaries (LMXBs, a binary where a compact star accretes mass from a Roche lobe filling companion star via an accretion disc) have so far been identified (see e.g. \citealt{Liu2007}) and about a third are proposed to harbour black holes (see \citealt{Corral-Santana2016,Tetarenko2016}). In spite of over a thousand papers dedicated to the topic over the past six decades, the evolution of an LMXB outburst (epochs where the brightness of the system increases dramatically over the course of a few weeks to years) remains a key outstanding question in the field. Previous works have established that the outburst properties are governed by two intrinsically connected processes, namely the accretion and ejection of mass (see \citealt{Fender2016} for a review). The two outflow signatures traditionally studied are radio jets, detected during the hard state (e.g., \citealt{Fender2004}), and X-ray winds, mostly detected during the soft state (e.g., \citealt{Neilsen2009,Ponti2012,Ponti2016,DiazTrigo2016}). More recently, recurrent outflows in the optical spectra of V404 Cygni during its 2015 outburst (\citealt{Munoz-Darias2016,MataSanchez2018}; hereafter MD16 and MS18, respectively; see also \citealt{Casares2019}; \citealt{Munoz-Darias2022}) were discovered. Their simultaneous detection with jet emission proved that these two mass ejection channels coexist in the  hard state. Since then, optical outflows have been detected in up to 9 LMXBs (see the most recent compilation in \citealt{Panizo2022}), effectively proving it is a common feature of the population. These low-ionisation (a.k.a. cold) winds are primarily detected during the hard and soft-intermediate states of the outburst (e.g., \citealt{Munoz-Darias2019}, hereafter MD19; also \citealt{MataSanchez2022}), though near-infrared observations suggest they might also be present during the soft state \citep{SanchezSierras2020}. The main tools employed for the identification of outflow features in optical and near-infrared spectra are limited to visual inspection or the so-called excesses diagnostic diagram (MS18). While the former typically relies on simultaneous detection of a number of features (including blue-shifted absorptions, extended emission wings, skewed profiles and flat-top lines) over different emission lines, the latter assumes a Gaussian-like underlying component from the accretion disc contribution at the wings of the profile (see also \citealt{Panizo2021}). However, as the population of systems with detected outflows continues to grow, it is required to develop automatic methods for the identification and classification of these features, in order to provide scalable and reproducible solutions.

Machine learning (ML) methods have revolutionised most scientific fields over the past decade. In astrophysics, these techniques have also been adopted, which proved to be particularly well suited to analyse large databases accumulated over years (e.g. \citealt{Ball2010, Huertas-Company2023}). The arguably most widespread use case is the automatic classification of vast observing samples, to the point that ML has become a standard part of large surveys discovery pipelines (e.g. \citealt{Brink2013,Wright2015,Killestein2021}). For the particular field of LMXBs, the application of these techniques has just started to permeate, being so far focused on their X-ray emission, attending to their spectral and timing properties (e.g., \citealt{Pattnaik2021,OrwatKapola2022,deBeurs2022,Ricketts2023}). 

The present work aims at constructing a neural network (NN) with the purpose of automatically classifying spectroscopic features traditionally associated with the presence of wind-type outflows. As a pilot study, this work provides a proof of concept on the methodology, technical feasibility, and limitations of the technique. We train the NN on a simulated database of spectral emission lines generated from theoretical disc models of LMXBs, to which we injected a variety of outflow features (Sec. \ref{sec:databases}). This allows us to test the performance of supervised ML methods on the simulated data (Sec. \ref{sec:method}), after which we apply the trained algorithm to observational data from outbursts of two different LMXBs (Sec. \ref{sec:realobs}). Comparison of the derived results with those obtained through traditional techniques enables us to determine if the ML approach is a convenient tool for the study of outflow spectral features.

\section{Databases}
\label{sec:databases}

This pilot study employs two types of databases, each serving to a different aim. The purpose of simulated databases is to train the supervised ML algorithms at the heart of the classification. On the other hand, observational databases enable us to test the trained classifiers on a real-case scenario, as well as to compare the outcome with traditional techniques dedicated to outflow detection. It follows a detailed description of each of these samples.

\subsection{Simulated databases}
\label{simdata}

\begin{figure*}
\includegraphics[keepaspectratio, trim=0cm 0cm 0cm 0cm, clip=true, width=\textwidth]{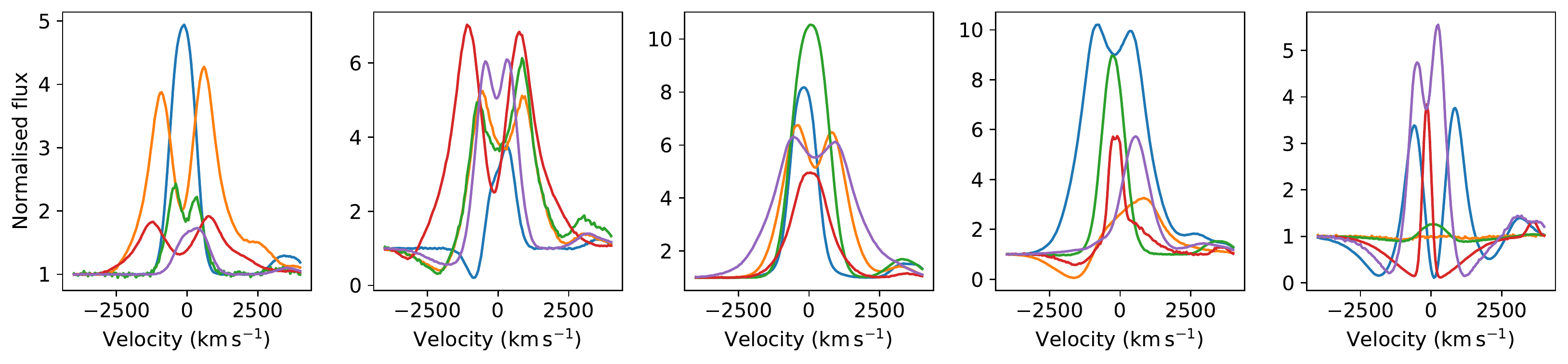}
\caption{Examples of the final, simulated spectra for the $\rm H\beta$ line, using the instrumental setup corresponding to GTC R2500R. Each panel corresponds to the five different classes defined in this pilot study. From left to right: \textit{disc}, \textit{blue-absorption}, \textit{broad}, \textit{P-Cygni} and \textit{absorbed}. Note that the emission line of He\textsc{i} 4922 (located at $\sim 3700 \, {\rm km\,s^{-1}}$) has been also included in the database, to better simulate the real observations.} \label{fig:Hbetaexample}
\end{figure*}

We construct a simulated database where we emulate emission line profiles arising in LMXB accretion discs, including typical outflow features identified in their observational spectra. Different components must be implemented in order to generate a realistic sample: an underlying disc profile reproducing the classic double-peaked emission line; the outflow features that we aim to detect and classify (such as P-Cygni profiles); and finally, line contaminants that might bias the ML performance (such as known interstellar/telluric lines, or emission lines from distinct but nearby species). We will create separate databases for the most frequently observed emission lines in the optical range of LMXBs: $\rm H\alpha$, $\rm He\textsc{i}\,5876$, $\rm H\beta$ and $\rm He\textsc{ii}\, 4686$. The simulated databases contain a million spectra each, equally split over five different classes: \textit{disc}, \textit{blue-absorption}, \textit{broad}, \textit{P-Cygni} and \textit{absorbed}. We provide below a detailed description of these classes and their associated parameters. Examples of generated line profiles for each class are displayed in Fig. \ref{fig:Hbetaexample} for the reader convenience.

\subsubsection{Disc emission line profile}
\label{baseprofile}

The optical spectra of LMXBs are dominated by the contribution from their accretion discs, which generates a broad emission line profile either single-peaked or double-peaked, depending on the binary parameters (e.g. orbital inclination) as well as the spectral resolution of the observations (defined by the particular instrumental setup).   

We employ the model described in equation 4 of \citet{Orosz1994}, adapted from \citet{Horne1986}, to generate the pure \textit{disc} emission line profile. We note that this model was originally developed to reproduce quiescent accretion discs, while the aim of the present project requires comparison with line profiles produced during the outburst (as optical outflows have been detected only during these brighter epochs). During the outburst, the accretion disc expands and heats up, becoming brighter while pushing the region responsible for the optical line formation to an outer (cooler) radius. Under the assumption of a Keplerian distribution of velocities for the gas in the disc, the velocity range of the line forming region during outburst should shrink, ultimately leading to narrower line profiles. This is considered the canonical profile evolution, and it has been observed in many LMXBs (e.g. Swift J1357.2-0933, \citealt{Torres2015,MataSanchez2015b,Corral-Santana2013}; MAXI J1820+070, MD19, \citealt{Torres2020}). Nevertheless, deviations from this scenario have also been reported, being the presence of outflows a proposed explanation for the widening of the line profile during outburst (e.g. V404 Cygni, MS18). For these reasons, we decided to employ the aforementioned models to generate our simulated database, but covering an adequate range of parameters to generate narrow enough profiles, comparable to those observed during outburst events. The limitations and biases introduced by this decision are further discussed in Sec. \ref{discussion}. 

The disc model presented here depends on a series of physical parameters for the binary system. To generate a prolific enough database in order to emulate all kinds of LMXB profiles, we define the following ranges for the different parameters at play:

\begin{itemize}
    \item $\alpha$: following \citet{Orosz1994}, we set the range for the exponent of the disc emissivity power-law to $1.5-1.6$. We note that the effect of this parameter on the line profile is small for the inspected range. 
    \item $r_1$: the innermost radius of the accretion disc, in units of the outer disc radius, is suggested to vary between $0.05-0.15$ \citep{Orosz1994}. This parameter limits the maximum velocities reached at the wings of the profile, and constrains the simulated profiles within a reasonable range. 
    \item $i$: the orbital inclination of the system is one of the key parameters controlling the width of the line profile. We allowed this parameter to explore the full range of possible values ($0-90\, \deg$), so as to generate both single- and double-peaked profiles.
    \item $v_{\rm d}$: the outermost disc velocity effectively sets the double peak separation. We allow it to vary between $300-1200\, \rm{km\,s^{-1}}$, a range that encompasses typical values for outburst spectra (\citealt{Orosz1994,Orosz1995}).
    \item $\beta$ and $\phi_{\rm orb}$: these two parameters allow one to emulate the effect of a hotspot in the data. They essentially modify the relative intensity between the red and blue peaks as a function of the orbital phase. We cover ranges of $\beta=0.0-0.2$ (up to a $20\%$ intensity ratio between the two peaks) and $\phi_{\rm orb}=0.0-1.0$ (i.e., all possible orbital phases).
\end{itemize}

The above parameters describe \textit{disc}, normalised spectra. However, to reproduce experimental spectra, a handful of additional parameters are required:

\begin{itemize}
    \item $\mu$: the centroid of the line profile, which allows us to simulate the systemic velocity. We cover the range $\pm 250\, {\rm km\, s^{-1}}$, consistent with the observed kick velocities in LMXBs \citep{Atri2019}.
    \item $h$: the height of the normalised line profile ultimately depends on the relative contribution between the emission line and the underlying continuum. As this ratio is known to change during the outburst (e.g., due to differing ionisation levels), we decided to model a range of heights consistent with those typically observed in LMXBs. In particular, depending on the modelled emission line, the maximum allowed height was different, being the largest value for H$\rm \alpha$ ($<20$) and the lowest for $\rm He\textsc{i}\, 5876$ ($<2.0$). The minimum value was defined in terms of the signal-to-noise ratio (SNR; see Sec. \ref{sec:noise}), imposing a minimum $>3\sigma_{\rm N}$ detection.
    \item $\rm FWHM_{broad}$: a number of effects, whether intrinsic or external to the LMXB, can lead to an additional broadening of the profile (characterised by its full-width-at-half-maximum, $\rm FWHM_{broad}$). The most straightforward is the instrumental setup, which inevitably introduces a broadening component in the observed spectrum due to the limited instrumental resolution (${\rm FWHM}_{\rm inst}$; we used either $200\, {\rm km\, s^{-1}}$ or $300\, {\rm km\, s^{-1}}$ for this pilot study, see Sec. \ref{obsdata}). Additionally, an intrinsic broadening component when comparing the disc models and observations has been recently reported \citep{Casares2022}. Even though the origin of this local broadening remains unclear, we simulate it by allowing for ${\rm FWHM}_{\rm local}$ values in the range $50 - 650 \, {\rm km\,s^{-1}}$ (following the results presented in \citealt{Casares2022}). We combine both effects by convolution of the disc spectrum with a Gaussian kernel of $\rm FWHM_{broad}=\sqrt{{\rm FWHM}_{\rm inst}^2+{\rm FWHM}_{\rm local}^2}$. 
\end{itemize}

Each combination of the above parameters generates a single emission line profile which is evaluated in the radial velocity space, defined with respect to the corresponding emission line rest wavelength. We decided to cover a range of $\pm 4000\,{\rm km\, s^{-1}}$ (which include the fastest outflow profiles ever detected, see MD16 and MS18), and impose a pixel size determined by the dispersion of the observational setup (either $50\, {\rm km\, s^{-1}}$ or $100\, {\rm km\, s^{-1}}$, see Sec. \ref{obsdata}).

We then use a Monte Carlo approach to generate, for a given observing setup, a database containing a million of simulated spectra. The parameter values employed to construct each spectrum were randomly drawn from uniform distributions within the aforementioned limits, except for the orbital inclination, for which a uniform distribution in $\cos{(i)}$ was employed instead (corresponding to an isotropic distribution of $i$). Out of the complete database generated and with the aim to define five balanced classes, $20\%$ will be kept untouched as pure \textit{disc} profiles, which will be classified as \textit{disc} type spectra. The remaining $80\%$ spectra will be modified to describe the four remaining classes detailed below.

\subsubsection{Outflow features}

The observational signatures of wind-type outflows can result in a wide variety of profiles in the optical spectra, depending on a number of conditions such as the outflow geometry, its density and temperature. Complex outflow models are out of the scope of this pilot study, which aims at detecting typical features generated by either optically thin or thick ejecta. As such, we focus on reproducing the most commonly observed profiles:

\begin{itemize}

    \item \textit{Blue-absorption}: an optically thick outflow launched towards the observer produces an absorption at blue-shifted wavelengths (with respect to that of the transition). We emulate this feature with a Gaussian of negative height (between $-50\%$ of the \textit{disc} profile normalised flux and the $3\sigma_{\rm N}$ detection level), with blue-shifted centroid (covering the range of $-500 $ to $ -2500 \, {\rm km\, s^{-1}}$), relative to the centroid of the \textit{disc} profile $\mu$), and $\sigma$ between that constrained by the spectral resolution and the $v_{\rm d}$ value (but always below $1000 \, {\rm km\, s^{-1}}$, to avoid unrealistically wide profiles).
    \item \textit{Broad}: an optically thin, isotropic and homogeneous nebula would produce instead an emission component across a broad range of projected velocities. This has been observed in a number of systems experiencing outflows, such as nova events (e.g., \citealt{Iijima2003}), and more recently in LMXBs outbursts (MD16). We implement this feature as a Gaussian in emission with height within the range between $3\sigma_{\rm N}$ detection and $4$ (up to $20$ for $\rm H\alpha$, as it has shown the most prominent examples); centroid fixed at the \textit{disc} profile $\mu$ value, and $\sigma$ limited to the range between $v_{\rm d}/2 $ and $v_{\rm d}/2+1000\, {\rm km\, s^{-1}}$ (up to $v_{\rm d}/2+1300\, {\rm km\, s^{-1}}$ for $\rm H\alpha$).
    \item \textit{P-Cygni}: the most robust outflow signatures are arguably P-Cygni profiles, present in a variety of systems (e.g., \citealt{Smith2006,Thone2017}). It is naturally produced by an homogeneous and isotropic expanding shell of matter, and produces a blue-shifted emission (due to the material ejected towards the observer) superimposed to a broad emission component (produced through line recombination over the entire shell). We will implement this profile as a combination of the two features described above (\textit{blue-absorption} and \textit{broad}), using the exact same range of values. 
\end{itemize}

These features are injected to the \textit{disc} profiles by randomly selecting parameter combinations from uniform distributions constrained within the previously defined limits. For each of the outflow classes defined above, we implement the corresponding outflow feature into $20\%$ of the simulated spectra (i.e., $60\%$ of the simulated database contains some kind of outflow signature).

\subsubsection{Broad absorption}

Broad absorption components have been observed in a number of outbursting LMXBs, such as Nova Muscae 91 \citep{dellaValle1991}, GRO J0422+32 \citep{Casares1995,Callanan1995,Shrader1997}, Nova Velorum 93 \citep{Masetti1997,dellavalle1997}, GRS J1655-40 \citep{Bianchini1997,Soria2000}, XTE J1118+480 \citep{Dubus2001}, MAXI J1807+132 \citep{JimenezIbarra2019b}, MAXI J1803$-$298 \citep{MataSanchez2022} and more recently MAXI J1348-630 \citep{Panizo2022}; to name a few. It can be described as a broad absorption where the emission line is embedded, with varying depth and width as the outburst evolves. Depending on its depth and centroid velocity (with respect to the emission line), it can give rise to a profile which might be confused with the \textit{blue-absorption} outflow case. The origin of this feature is still under debate, with the most accepted explanation invoking self-absorption by the accretion disc atmosphere \citep{Dubus2001}. The automatic identification of this feature will be a key tool to study its origin through future population studies, but it is also required in the present paper to avoid flagging it as an outflow. We simulate it as a low-velocity, broad absorption Gaussian with centroid in the range of $\mu\pm 400\, {\rm km\, s^{-1}}$, height varying between $-50\%$ of the \textit{disc} profile normalised flux at the absorption centroid and the $3\sigma_{\rm N}$ detection level, and $\sigma$ between $2v_{\rm d}$ (with an absolute minimum of $300\, {\rm km\, s^{-1}}$) and $1100\, {\rm km\, s^{-1}}$ (up to $1300\, {\rm km\, s^{-1}}$ for $\rm H\alpha$). We apply such broad absorption feature to the remaining $20\%$ of the generated spectra.

\subsubsection{Line contaminants}

Real spectral lines of LMXBs might exhibit other features close to the transition of interest that effectively contaminate the observations. In this work, we decided to simulate separate databases for the following emission lines: $\rm H\alpha$, $\rm He\textsc{i}\,5876$, $\rm H\beta$ and $\rm He\textsc{ii}\, 4686$. The main contaminants for each of these lines, as well as the process to include them into the simulated databases, are as follows:

\begin{itemize}
    \item $\rm H\alpha$: a telluric line is present at $6613$\AA, which could influence the profile of particularly broad $\rm H\alpha$ lines. For this reason, we simulated the contaminant as a Gaussian absorption component at the corresponding wavelength, $\sigma$ defined by the spectral resolution, and height randomly picked from a uniform distribution between $-0.1$ and $0.0$.  
    \item $\rm He\textsc{i}\,5876$: the main contaminant to this line is the Na doublet at $5890$ \AA, a pair of absorption lines with a mixed origin (telluric and interstellar). We simulate them as two Gaussians centred at the corresponding wavelengths, with $\sigma$ determined by the spectral resolution, and height randomly picked from a uniform distribution between $-0.4$ and $-0.2$.
    \item $\rm H\beta$: the He\textsc{i} line at 4921.929 \AA\,, even if typically weaker, is close enough to $\rm H\beta$ to influence its profile on certain spectra. Therefore, we simulate it as an accretion disc line with the exact same model parameters employed for the $\rm H\beta$ line, but allowing for slightly wider peak separation (randomly picked from a uniform distribution between $v_{\rm d}$ and $1.5\,v_{\rm d}$), due to the line being produced at a slightly different radius, and smaller height ($0.07-0.2$ times that of $\rm H\beta$).
    \item $\rm He\textsc{ii}\, 4686$: this is the trickiest line of our sample, due to the presence of both the $\rm He\textsc{i}\, 4713$ line, and the Bowen blend at $\sim 4640$ \AA\,. The former is typically weak, and we model it as an accretion disc line with the same parameters as $\rm He\textsc{ii}\, 4686$, but height within the 0.05-0.25 range. The later is a mixture of C \textsc{iii} and N \textsc{iii} emission lines, which appears blended into a single broad component close to $\rm He\textsc{ii}\, 4686$ and of comparable intensity. We attempt to simulate this complicated profile as a broad Gaussian emission line, with $\sigma$ between $2000-2700 \,{\rm km\, s^{-1}}$, and height of $0.1-0.55$ times that of $\rm He\textsc{ii}\, 4686$.
\end{itemize}

\subsubsection{Simulated noise}
\label{sec:noise}
In order to create a more realistic database, we decided to add noise to the simulated spectra. We implemented it by using a single parameter: SNR. For a given spectrum, at a particular velocity pixel, we sample a value from a uniform distribution in the range of $\rm SNR=30-330$, and we create a Gaussian distribution centred at the normalised spectrum flux value ($f_{\rm norm}$) with a standard deviation of $\sigma_{\rm N} = \sqrt{f_{\rm norm}}/{\rm SNR}$. We randomly draw samples from this distribution, which are added on top of the original $f_{\rm norm}$ of the \textit{disc} spectrum. Note that this approximation assumes a constant underlying continuum under the emission line, a reasonable assumption for such a small wavelength range. It also produces noise with $\sigma_{\rm N} \sim 1/{\rm SNR}$ at the normalised continuum level, while improving on the relative uncertainties at the peak of the emission line. 

\subsubsection{A zoo of profiles}

\begin{figure*}
\includegraphics[keepaspectratio, trim=0cm 0cm 0cm 0cm, clip=true, width=\textwidth]{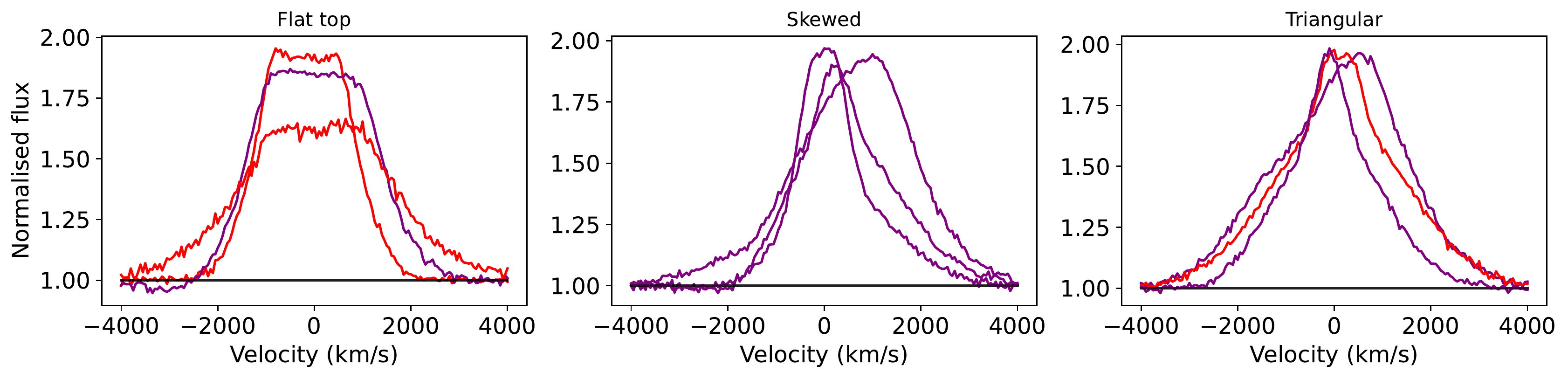}
\caption{Simulated line profiles randomly generated with peculiar shapes reminiscent of more complex outflow-related profiles. Red colour corresponds to \textit{broad} type and purple to \textit{P-Cygni} type simulated spectra.} \label{fig:peculiarrofiles}
\end{figure*}

During the generation of the simulated database, a number of the randomly generated line profiles attracted our interest. The most conspicuous among them are shown in Fig. \ref{fig:peculiarrofiles}, and include flat-top, triangular and skewed emission lines. These have been previously observed in LMXBs and associated to the presence of outflows (e.g. \citealt{SanchezSierras2020}, \citealt{Panizo2022}). We are aware that similar profiles have also been observed in other systems with outflows but without accretion discs (e.g. supernovae, see \citealt{Smith2014}), which implies that the double-peaked component is not mandatory to produce the aforementioned outflow signatures. Indeed, these can arise as a result of different combinations of outflows geometries and optical thickness. Nevertheless, this also implies that our simulated database, in spite of not employing a physical model to simulate the outflow components, is already able to produce such profiles. While describing the physical formation mechanism of the outflow profiles is beyond the scope of this paper, we wanted to highlight that a simple combination of no more than two Gaussians (respectively in absorption and emission) on top of the double-peaked \textit{disc} profile is already able to naturally produce these characteristic shapes, and therefore, they will be considered during the algorithm training.

\subsection{Observational datasets}
\label{obsdata}

In order to test the ML algorithm trained on the simulated database, comparison with real, observed normalised spectra from LMXBs is required. For this pilot study, we decided to focus on two systems for which particularly prolific databases were compiled during their latest outbursts.

\begin{itemize}
    \item V404 Cygni: this LMXB, which contains the first dynamically confirmed stellar-mass black hole \citep{Casares1992}, exhibited an extremely bright outburst in 2015 after $\sim 25$ years of quiescence \citep{Barthelmy2015}. Intensive multi-wavelength follow up of the event led to dozens of publications (e.g. \citealt{Kimura2016,Motta2017}). Of particular relevance for this work is the first systematic analysis of the appearance and evolution of optical outflows in a LMXB (MD16), which sparked the interest of the community in such features and their influence in the outburst evolution. In this pilot study, we will limit our analysis to a sub-sample of the database presented in MS18. In particular, we only consider the 510 spectra obtained with the Gran Telescopio Canarias (GTC) and grism R1000B, which cover the first 15 days of the outburst. 
    \item MAXI J1820+070 (herefater MAXI J1820): the discovery outburst of this LMXB \citep{Tucker2018} was also followed with great interest, due to both its brightness and accessibility from all over the globe. The numerous studies dedicated to this event allowed to unveil both a high orbital inclination and a black hole nature for the compact object (see e.g., \citealt{Torres2019,Torres2020}). The detection of outflow features in both optical and near-infrared spectra has also been reported (MD19, \citealt{SanchezSierras2020}), and a significant database of spectra has been collected. In this pilot study, we focus on the subsample of 48 spectra observed with GTC and grism R2500R presented in MD19, which cover both the hard and soft state of the outburst spanning over eight months.
\end{itemize}

Both databases have been previously described thoroughly in dedicated studies to each object, so we refer the reader to the aforementioned papers in order to learn the data reduction details. 

\subsection{Databases homogenization}

Direct comparison of the simulated and observational databases is not straightforward. To this end, we extracted cuts of the observed spectra centred at the lines of interest: $\rm H\alpha$, $\rm He\textsc{i}\, 5876$, $\rm H\beta$ and $\rm He\textsc{ii}\, 4686$. We binned all the spectra in velocity space to a uniform pixel size consistent with the dispersion of each instrument ($50\, {\rm km\, s^{-1}}$ and $100\, {\rm km\, s^{-1}}$ for R2500R and R1000B, respectively) and covering the range of $\pm 4000\,{\rm km\, s^{-1}}$. We then created separate, simulated databases using the corresponding dispersion values mentioned above and the associated spectral resolution with each grism configuration (${\rm FWHM}_{\rm inst}=200\, {\rm km\, s^{-1}}$ and $300\, {\rm km\, s^{-1}}$ for R2500R and R1000B, respectively).

\section{Methods and results}
\label{sec:method}
The aim of this work is to apply ML techniques to identify outflow features in LMXBs emission lines. Thanks to the explosive growth of the ML field and its application across different disciplines, new tools have been developed, tested and made available to the community. All the ML classifiers implemented in this work belong to the NN type and have been developed making use of the open source deep learning library \textsc{keras} \citep{Chollet2015} with the \textsc{Tensorflow} \citep{Abadi2015} back-end.

\subsection{NNs classifiers}

We decided to focus on Deep Neural Networks (DNNs), which since their original success for solving computer vision problems \citep{Krizhevsky2012} have been applied to an extensive range of cases (see e.g., \citealt{Szegedy2015}). DNNs performance seems to particularly excel on speech recognition (e.g., \citealt{Sainath2013}) and natural language processing tasks (e.g., \citealt{Mikolov2013}), both of which are characterised by their sequential nature, also a common property for time series and spectra. 

We explore three types of supervised classifiers of the DNN class, aiming to classify our simulated database into the five different classes defined in Sec. \ref{simdata}. We provide below a short description of the parameters defining the different layers compounding each algorithm, while referring the reader to \citet{Fawaz2018} and references therein for an in-depth description of the algorithms themselves:
 
\begin{itemize}
    \item Multi Layer Perceptrons (MLP): this network treats independently time series/spectral elements from each other, which means the sequential aspect of the data is ignored. We constructed a network containing 4 dense layers whose outputs are fully connected. They are three hidden layers of 500 neurons and Rectified Linear Unit (ReLU) activation function, and a final softmax classifier. Each layer is preceded by a dropout \citep{Srivastava2014} with rates of 0.1, 0.2, 0.2 and 0.3, respectively, a form of regularization preventing overfitting.
    
    \item Fully Convolutional Neural Network (FCN): a type of convolutional neural network, proposed in \citet{Wang2017b} and composed of three convolutional blocks. Each consists of a convolution, followed by a batch normalization (\citealt{Ioffe2015}; a normalization layer to help the network converge quickly), and a ReLU activation function. The three convolution layers have 128, 256 and 128 filters; with respective lengths of 8, 5 and 3.
    
    \item Residual Network (ResNet): this architecture, also proposed by \citet{Wang2017b}, consists of three residual blocks. Each of them is composed by three convolution layers with a ReLU activation function and preceded by a batch normalization, whose output is fed to the input of the residual block. The convolutions within a residual block all have the same number of filters (which is 64, 128 or 128, for each different residual block), and different filter lengths (8, 5 and 3; for each set of convolution layers within the block).
\end{itemize}

\subsection{Training and performance assessment}
\label{trainingandperformance}

\begin{figure}
\includegraphics[keepaspectratio, trim=0cm 0cm 0cm 0cm, clip=true, width=0.5\textwidth]{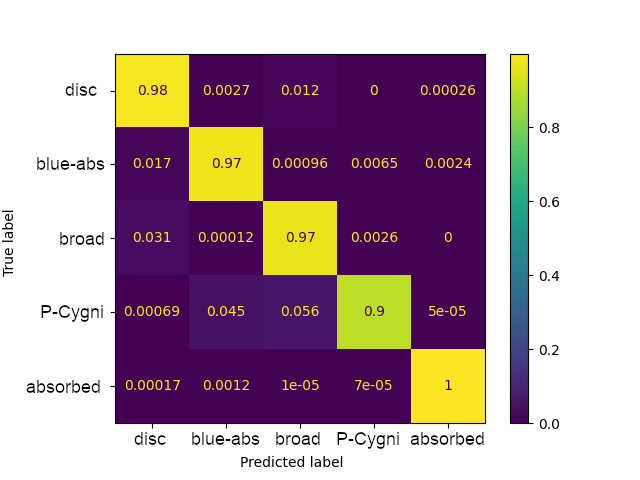}
\caption{Confusion matrix for the ResNet classifier when applied to the $\rm H_\alpha$ R2500R database. Each cell value has been normalised by the total number of spectra per class}  \label{fig:confmat}
\end{figure}

We independently trained the three algorithms described above on the simulated spectral databases previously introduced. This implies the independent training of six databases: four for the grism setup R1000B (one per line profile: $\rm H\alpha$, $\rm He\textsc{i}\, 5876$, $\rm H\beta$ and $\rm He\textsc{ii}\, 4686$) and two for the R2500R ($\rm H\alpha$, $\rm He\textsc{i}\, 5876$). As part of the training, we evenly split each database into training and test samples. We flagged $30\%$ of the training dataset as a validation dataset in order to assess the training performance. We compiled the models with an \textsc{adam} optimizer and a $\textsc{categorical crossentropy}$ loss function. We set an early stopping condition defined by the lack of improvements over the last 50 epochs on the loss function, measured in the validation set. The performance is then evaluated by applying the trained classifier to the unseen test database, which produces the metrics reported in Table \ref{tab:metrics} (see also an example of confusion matrix in Fig. \ref{fig:confmat}). We note that the actual output of the NN is an array containing the probability for each particular spectrum to be associated with each of the five classes (as determined by the final, softmax layer). We assign to each processed spectrum the class with the highest probability for the metric comparison purposes described above. Inspection of the individual Receiver Operating Characteristic (ROC) curves for each class and setup reveals area under the curve (AUC) values higher than $0.98$. We also explored the thresholds on the classification probability based on the ROC curves that optimize the true positive rate (tpr) and the false positive rate (fpr), and derived individual results for each setup, always keeping $\rm{tpr}>0.93$ and $\rm{fpr}<0.07$. The resulting threshold values are comparable between line types and setups, but vary significantly between the different classes, with 0.5 (\textit{absorbed}) and 0.1 (\textit{P-Cygni}) being at the extremes, coinciding with the best and worst performing classes, respectively. To simplify our selection criteria, we will hereafter keep the most restrictive threshold value of 0.5. Classifications below this threshold should be taken with caution, and when possible, independently confirmed.

The comparison of the performance of the different methods (attending only to their average metrics) reveals that the MLP script is producing the less accurate classification, down to $0.78$ for some databases. The average accuracy of both the FCN and ResNet algorithms is significantly better ($0.89-0.97$), with a slightly better performance favouring the ResNet case ($0.92-0.97$). A closer inspection of each individual confusion matrix shows that the main bulk of miss-classified spectra actually corresponds to blends of outflow classes (i.e., \textit{blue-absorption}, \textit{broad} and \textit{P-Cygni} class). Spectra from the \textit{P-Cygni} type leak into the \textit{blue-absorption} and \textit{broad} classes (with typical recall values for the \textit{P-Cygni} class of $\sim 0.85-0.90$), or alternatively, \textit{blue-absorption} and \textit{broad} class spectra are classified as \textit{P-Cygni} (with recalls of $\sim 0.90-0.97$ and $\sim 0.92-0.97$ for the \textit{blue-absorption} and \textit{broad} classes, respectively). This is not unexpected, as the \textit{P-Cygni} class is generated as a combination of profiles from the two other outflow categories. Therefore, if one of the two components (either the blue-shifted absorption or the broad wing emission) is too shallow (we simulate outflow features with depth/high over the $>\rm 3\sigma$ threshold defined by the SNR), the algorithm might not be able to classify them correctly. On the other hand, the \textit{disc} and \textit{absorbed} classes show a much more consistent performance, with stable recall values of $>0.96$.

To further investigate the origin of these miss-classifications, we produced histograms of the parameters employed during the database simulation, and compared them with those incorrectly classified. This allows us to inspect possible caveats in the parameters range determination when creating the database, as well as to explore the limitations of the ML classifiers. We conclude that their accuracy is mainly affected by the following cases:

\begin{itemize}

    \item Low velocity and shallow P-Cygni profiles: the low velocity P-Cygni profiles ($\lesssim 1000\, {\rm km\, s^{-1}}$), specially if they are also shallow, are harder to identify as they fall between the peaks of the \textit{disc} emission line profile. Rather than a clear absorption, this effectively creates an asymmetry in the line profile which might be confused with those produced by a hotspot. 
    \item Low height and narrow emission wings: the detection of a shallow broad wing component is specially challenging if the outflow components are relatively narrow (${\rm \sigma}\lesssim 500\, {\rm km\, s^{-1}}$). In such a case, only the central region of the \textit{disc} profile is altered, an effect that can be mimicked through slight modifications of the \textit{disc} profile parameters.
    \item Low SNR: the detection of features in noisy spectra is a challenging task, but it does not seem to be a critical parameter for the range considered ($30-330$), as the number of miss-classified spectra between the extremes does not vary drastically (typically a factor $< 3$). It is only the dominant factor for the \textit{disc} class spectra.
\end{itemize}

All of the above are expected limitations not exclusive to the ML approach. On this regard, it is worth remarking that the ML technique does not appear to introduce any new biases, other than those inherent to the data quality and particular outflow properties.

    \begin{table}
	\caption{Performance metrics for the MLP, FCN and ResNet classifiers evaluated over the test dataset, corresponding to the averaged values over all classes.}	
	\begin{tabular}{l c c c c l}
             & &   \textbf{MLP}& & \\
		\hline
		 Line & Grism  & Precision & Recall & $\rm F_1$ & Accuracy   \\        
		\hline
        \hline
          $\rm H\alpha$ & R1000B & 0.85 & 0.82 & 0.81 & 0.82 \\
           & R2500R & 0.81 & 0.78 & 0.77 & 0.78 \\
          $\rm He\textsc{i}\, 5876$ & R1000B & 0.91 & 0.90 & 0.90 & 0.90\\
           & R2500R & 0.88 & 0.87 & 0.87 & 0.87 \\
          $\rm H\beta$ & R1000B & 0.90 & 0.88 &0.88 & 0.88 \\
          $\rm He\textsc{ii}\, 4686$ & R1000B & 0.83 & 0.81 & 0.82 & 0.81 \\         
 		\hline
        \\
                & &   \textbf{FCN}& & \\
 		\hline
		 Line & Grism  & Precision & Recall & $\rm F_1$ & Accuracy   \\        
        \hline\hline
          $\rm H\alpha$ & R1000B & 0.95 & 0.95 & 0.95 & 0.95 \\
           & R2500R & 0.96 & 0.96 & 0.96 & 0.96 \\
          $\rm He\textsc{i}\, 5876$ & R1000B & 0.94 & 0.94  & 0.94 & 0.94 \\
           & R2500R & 0.95 & 0.95 & 0.95 & 0.95 \\
          $\rm H\beta$ & R1000B & 0.92 & 0.92 & 0.92 & 0.92 \\
          $\rm He\textsc{ii}\, 4686$ & R1000B & 0.89 & 0.89 & 0.89 & 0.89 \\   
 		\hline
        \\
                  & & \textbf{ResNet}& & \\
 		\hline
		 Line & Grism  & Precision & Recall & $\rm F_1$ & Accuracy   \\        
        \hline\hline
          $\rm H\alpha$ & R1000B & 0.95 & 0.95 & 0.95 & 0.95 \\
           & R2500R & 0.97 & 0.96 & 0.96 & 0.96 \\
          $\rm He\textsc{i}\, 5876$ & R1000B & 0.95 & 0.94 & 0.94 & 0.94 \\
           & R2500R & 0.96 & 0.96 & 0.96 & 0.96 \\
          $\rm H\beta$ & R1000B & 0.94 & 0.93 & 0.93 & 0.93\\
          $\rm He\textsc{ii}\, 4686$ & R1000B & 0.93 & 0.92 & 0.92 & 0.92\\   
		\hline
    \end{tabular}
    \label{tab:metrics}
    \end{table}

\section{Application to observational spectra}
\label{sec:realobs}

After assessing the performance of the different NNs on the test and validation datasets, we select ResNet as the most efficient among them. Further discussion will be limited to results produced by this classifier, which will be hereafter referred to as the Ask The Machine (ATM) algorithm.

We applied the trained ATM classifiers to the real spectroscopic datasets described in Sec. \ref{obsdata}. For the V404 Cygni dataset, this includes four classifiers dedicated to each of the lines detected with the R1000B grism setup, while for MAXI J1820 data it only includes the two lines covered by the R2500R grism. A summary of the results from each classification is collected in Table \ref{tab:realclass}, but also discussed below for each dataset and line separately.

We will compare the ATM classification with the two most widely used traditional techniques for outflow detection: the excesses diagnostic diagram and the visual inspection. The excesses diagnostic diagram was first introduced in MS18 and further refined in \citealt{Panizo2021}. It is based on the assumption that departures in the wings of the inspected profile from a Gaussian function are due to the presence of outflows. After a Gaussian fit is subtracted from the line profile and the core of the line masked (on a case-by-case basis), equivalent widths (EWs) are measured on the residuals at regions corresponding to the red and blue wings of the profile ($\rm{EW}_r$ and $\rm{EW}_b$, respectively). The position of a particular spectrum in the excesses diagnostic diagram ($\rm{EW}_b$ against $\rm{EW}_r$) determines the presence and class of the outflow feature. On the other hand, detection of outflow signatures through visual inspection typically requires the presence of simultaneous features associated with outflows in different lines.

    \begin{table*}
	\caption{Results from applying the ATM classifier to real databases, where we report the number of spectra belonging to each class. Note that for each combination of line and grism, ATM was trained with a slightly different simulated database in order to better reproduce the corresponding setup conditions. We also report within parentheses those spectra whose classification is uncertain, as it falls below the 0.5 threshold described in Sec. \ref{trainingandperformance}.}	
	\begin{tabular}{l c c c c c c l}
		\hline
		 Database & Line & \textit{Disc}  & \textit{Blue-absorption} & \textit{Broad} & \textit{P-Cygni} & \textit{Absorbed}   \\        
		\hline\hline 
  
            V404 Cygni & $\rm H\alpha$ & 6 (4) & 0 (0) & 235 (40) & 269 (4) & 0 \\
           & $\rm He\textsc{i}\, 5876$ & 203 (4) & 193 (6) & 48 (1) & 65 (2) & 1 (0) \\
           & $\rm H\beta$ & 246 (0) & 80 (2) & 87 (3) & 87 (4) & 10 (0) \\
           & $\rm He\textsc{ii}\, 4686$ & 67 (5) & 268 (3) & 51 (6) & 65 (3) & 59 (3)  \\

		\hline
              MAXI J1820 & $\rm H\alpha$ & 36 (0) & 0 (0) & 12 (0) & 0 (0) & 0 (0) \\
           & $\rm He\textsc{i}\, 5876$ & 28 (1) & 19 (1) & 0 (0) & 0 (0) & 1 (0)  \\ 

		\hline
	\end{tabular}
    \label{tab:realclass}
    \end{table*}

\begin{figure*}
\includegraphics[keepaspectratio, trim=0cm 0cm 0cm 0cm, clip=true, width=\textwidth]{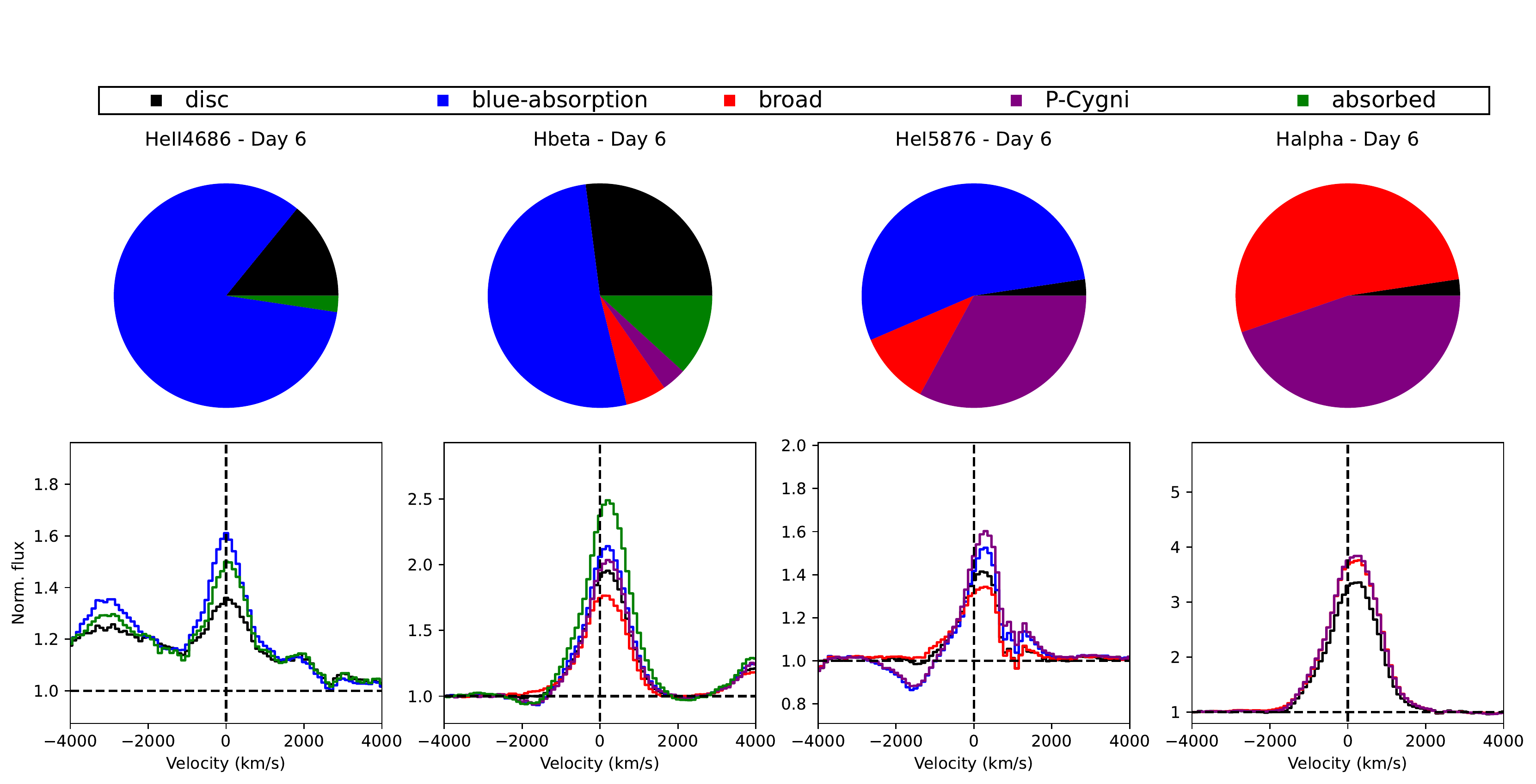}
\caption{Top row: pie chart representation of the classification with ATM for V404 Cygni spectra observed during day 6, where each colour correspond to a different class as described in the legend. Bottom row: the resulting averaged spectrum from the combination of all the observed spectra during this run, separated by the identified classes and following the same colour code.}  \label{fig:realclassexample}
\end{figure*}

\subsection{V404 Cygni}

\begin{figure*}
\includegraphics[keepaspectratio, trim=0cm 0cm 0cm 0cm, clip=true, width=0.5\textwidth]{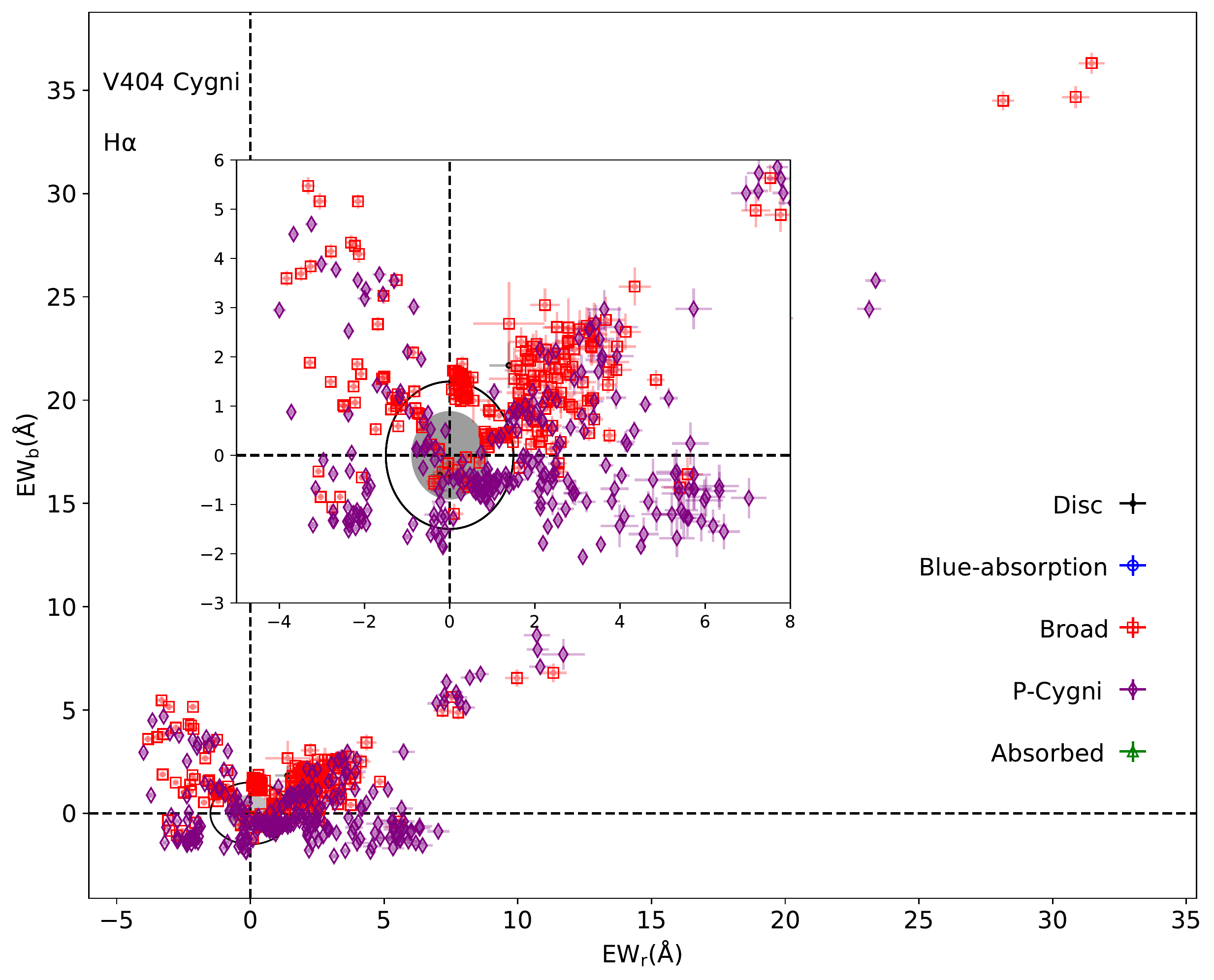}\includegraphics[keepaspectratio, trim=0cm 0cm 0cm 0cm, clip=true, width=0.5\textwidth]{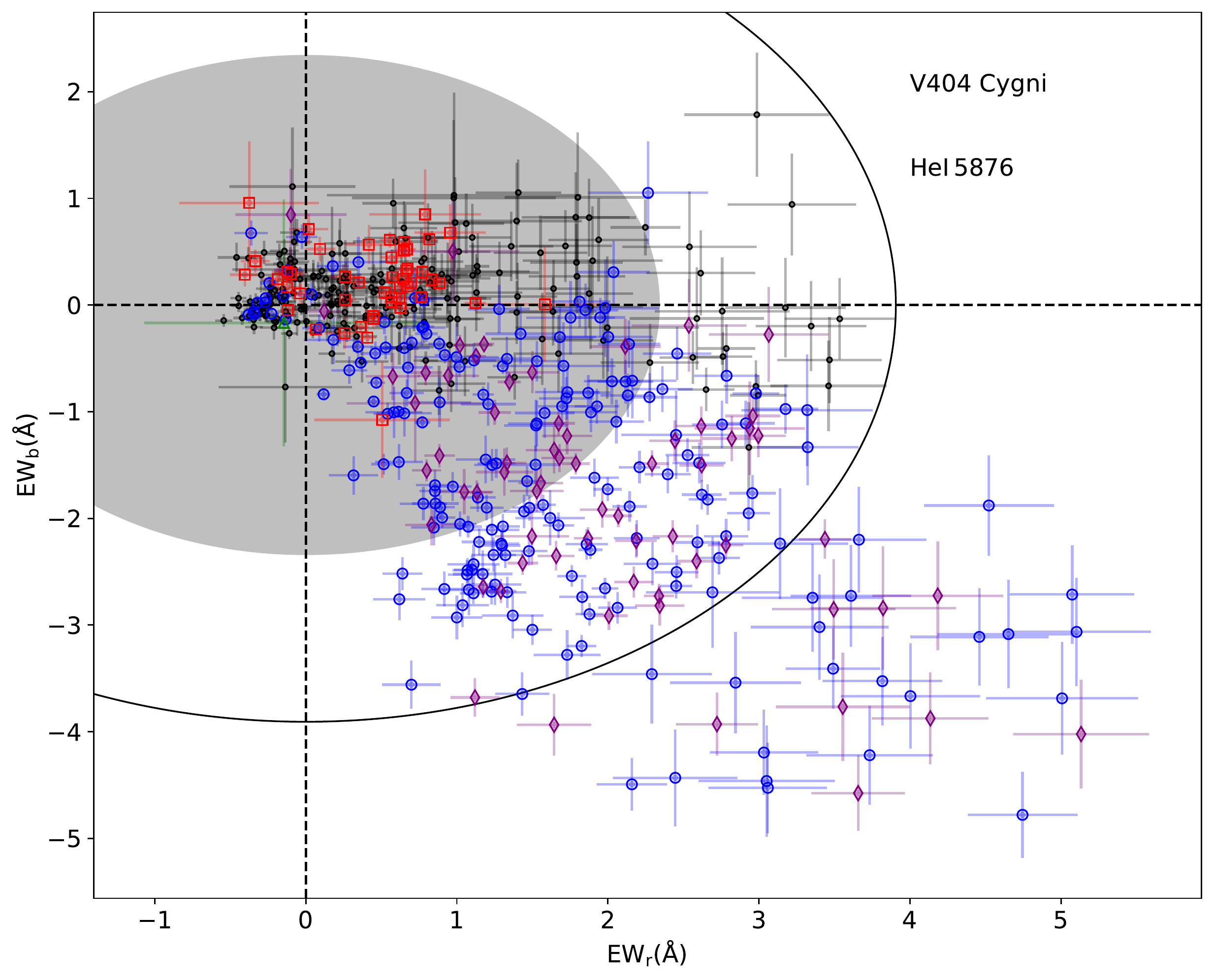}
\includegraphics[keepaspectratio, trim=0cm 0cm 0cm 0cm, clip=true, width=0.5\textwidth]{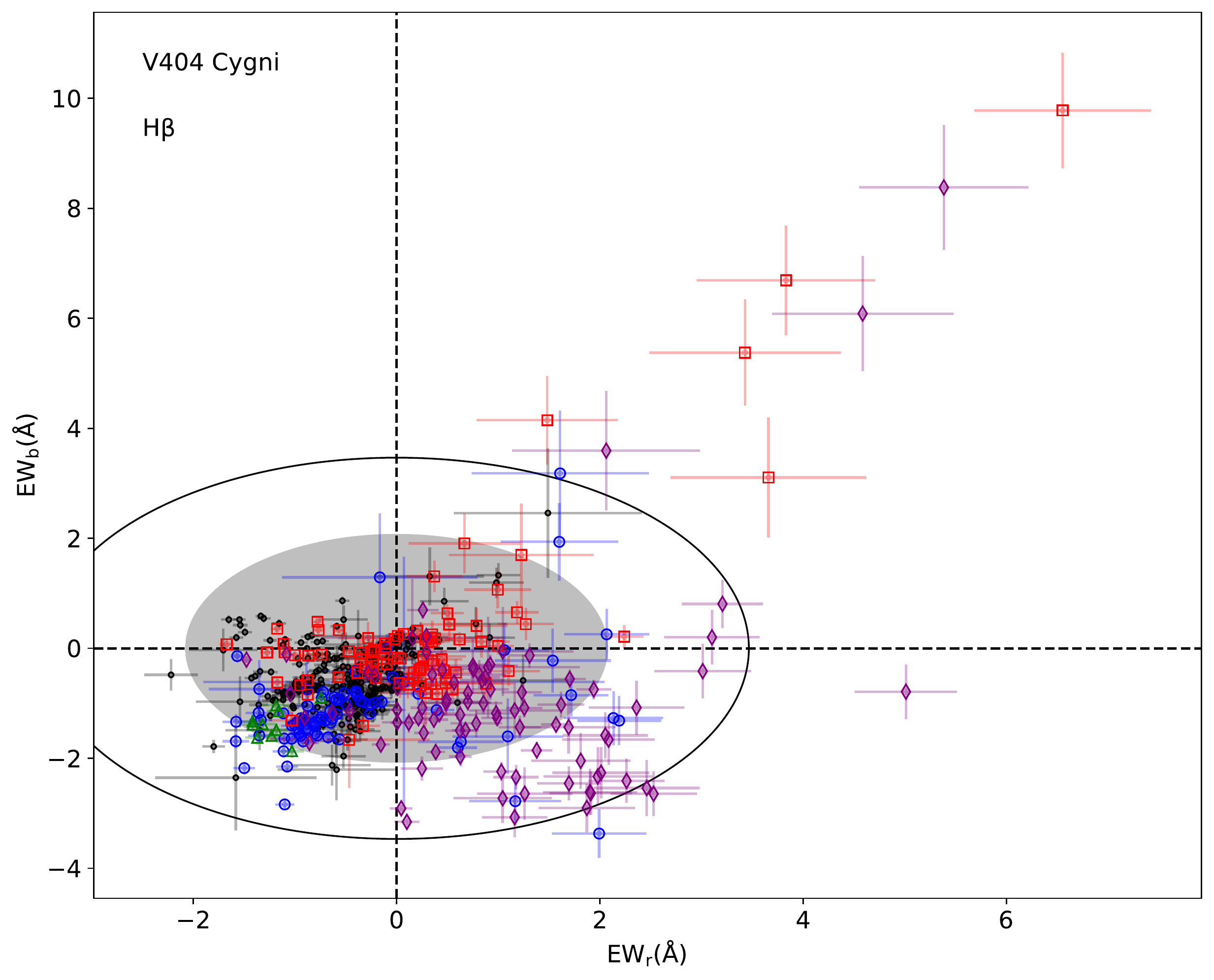}\includegraphics[keepaspectratio, trim=0cm 0cm 0cm 0cm, clip=true, width=0.5\textwidth]{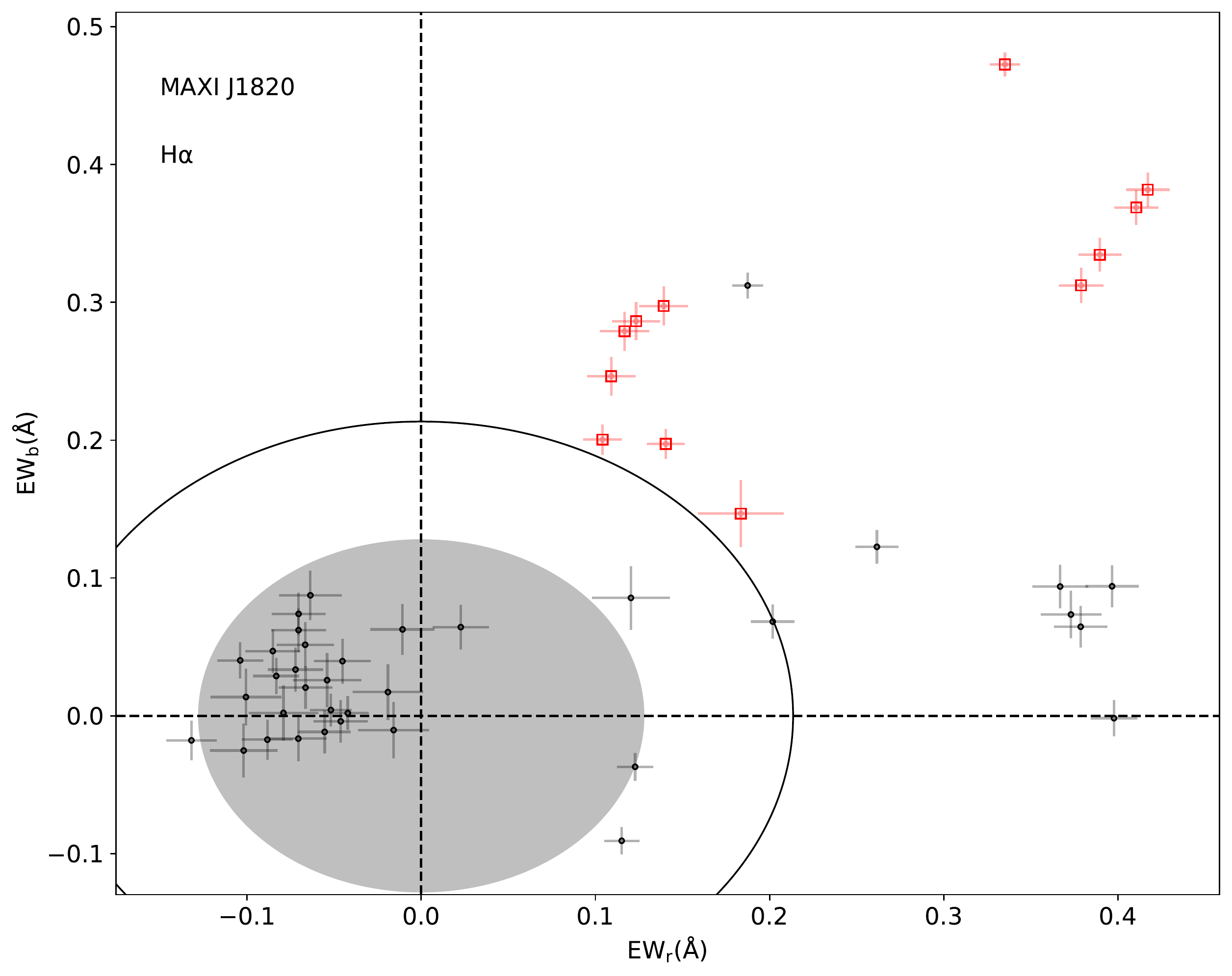}
\caption{Excesses diagnostic diagram applied to the lines of interest in each of the individual spectra from V404 Cygni and J1820 observational datasets. We do not show $\rm He\textsc{i}\, 4686$ for V404 Cygni neither $\rm H\beta$ for MAXI J1820 due to their line contaminants, which prevent a reliable application of the traditional method. The colour code matches that of Fig. \ref{fig:realclassexample}, depicting the ATM classification for each of the spectra: \textit{disc} (black dots), \textit{blue-absorption} (blue empty circles), \textit{broad} (red empty squares), \textit{P-Cygni} (purple empty diamonds) and \textit{absorbed} (green empty triangles) types. V404 Cygni correspond to the top panels, from left to right, $\rm H\alpha$ ($(\pm) 1000-4000\, {\rm km\, s^{-1}}$; including an inset of the central region) and $\rm He\textsc{i}\, 5876$ ($-4000 $ to $-500\, {\rm km\, s^{-1}}$; $1250 $ to $ 4000\, {\rm km\, s^{-1}}$; asymmetric due to the $\rm Na\textsc{i}$ interstellar doublet). Bottom left panel also corresponds to V404 Cygni, but focused on the profile of $\rm H\beta$ ($(\pm) 1000-2500\, {\rm km\, s^{-1}}$). Bottom right panel shows the MAXI J1820 excess diagram for the $\rm H\alpha$ line ($(\pm) 1000-4000\, {\rm km\, s^{-1}}$). All diagrams also include the $3\sigma$ (shaded) and $5\sigma$ (solid line) ellipses defining the excess diagram detection limits.}\label{fig:dexcess}
\end{figure*}

The large database collected during the 2015 outburst of V404 Cygni is extremely diverse, showing a wide range of profiles for each of the inspected lines. We report in Fig. \ref{fig:realclassexample} the results from applying ATM to a particular epoch (day 6, compound of 85 individual spectra and showing a large variety of outflow profiles) as a visual example of the classification results. We also exhibit the complete excesses diagnostic diagrams for the entire V404 Cygni database on $\rm H\alpha$, $\rm He\textsc{i}\, 5876$ and $\rm H\beta$  (see Fig. \ref{fig:dexcess}). In addition, we include their ATM classification in order to compare both techniques. It follows a detailed description for each line.

\subsubsection{$\rm H\alpha$}

Out of the complete sample consisting in 510 spectra, all but six are classified by ATM as having outflows in $\rm H\alpha$ (see Table \ref{tab:realclass}). In particular, the algorithm detects outflows of two types: either \textit{broad} ($46\%$) or \textit{P-Cygni} ($53\%$) profiles. The scarce number of spectra without outflow features is shocking, as such features are typically elusive in most LMXBs outburst events. Indeed, these outflows have only been detected systematically during the past few years through dedicated observations with the largest available telescopes. At this point, it is worth considering that a flawed design/training of ATM classification algorithm might be at play, biasing it to flag outflows in most spectra. However, it will become clear in the following sections that the ATM script is indeed able to identify \textit{disc} profiles (without outflows) in other lines and targets. Furthermore, previous works on the 2015 outburst of V404 Cygni (MD16, MS18, \citealt{Casares2019}) have shown the dramatic evolution of the $\rm H\alpha$ line profile, including extreme EWs, the detection of high-velocity outflows and a FWHM above the quiescence level during most of the event (opposite to the canonical behaviour, see Sec. \ref{baseprofile}). For these reasons, we believe the results from the ATM classification to be correct, and instead reveal V404 Cygni outburst as an extraordinarily rich event where outflows are always present. As a matter of fact, X-ray observations contemporaneous to those used here also show the presence of X-ray winds with similar observational properties to the optical ones (\citealt{Munoz-Darias2022}; see also \citealt{King2015})

Direct comparison of the results of the ATM with those drawn in MS18 from the excesses diagnostic diagram shows that this technique also found numerous P-Cygni like and nebular phase like profiles (\textit{P-Cygni} and \textit{broad} classes in this work, respectively). We show an updated version (following \citealt{Panizo2021} prescription) of the $\rm H\alpha$ excesses diagnostic diagram in Fig. \ref{fig:dexcess} (top left panel), which remains consistent with the original presented in MS18. This reveals an overall agreement with the ATM results: \textit{broad} spectra are mainly clustered on the top-right region of the diagram, while \textit{P-Cygni} type spectra dominate the bottom-right region (i.e., the P-Cygni region). It is worth remarking that the excesses diagnostic diagram only evaluates the wings of the line by comparing them with a Gaussian model, while it disregards asymmetries of the core of the line (which is effectively masked). Therefore, it is not surprising to find spectra in the nebular phase region being classified by ATM as \textit{P-Cygni} type: this suggests that, while broad wings are probably dominant at the edge of the profile, blue-shifted absorptions are also required to reproduce the observations. Another limitation of the traditional excesses diagnostic diagram, already highlighted in MS18, was the detection of apparent inverse P-Cygni profiles on day 6 (top-left region of the diagram). Visual inspection revealed they are actually caused by the presence of the intense P-Cygni profiles of the nearby $\rm He\textsc{i}\, 6678$ line, as well as a clearly asymmetric $\rm H\alpha$ line. On the other hand, ATM directly classifies day 6 spectra as either \textit{broad} or \textit{P-Cygni} types (see Fig. \ref{fig:realclassexample}), with the sole exception of two spectra that are classified as \textit{disc} at a probability below the threshold level of 0.5 (see Sec. \ref{trainingandperformance}). This shows the ability of ATM to overcome previous classification issues.

\subsubsection{$\rm He\textsc{i}\, 5876$}

ATM classification of this line profile reveals the following distribution among classes: $40\%$ \textit{disc} spectra, $60\%$ with outflow features (distributed as $63\%$ \textit{blue-absorption}, $16\%$ \textit{broad} and $21\%$ \textit{P-Cygni} profiles) and a single spectra of the \textit{absorbed} type. The detection of a significant number of \textit{disc} profiles is reassuring, supporting the aforementioned argument that the continuous presence of outflows in $\rm H\alpha$ is real, rather than derived from fundamental issues in ATM design and training. Most of the profiles classified as outflows contain a blue-shifted absorption, sometimes accompanied by a broad emission component. This is consistent with the results of the excesses diagnostic diagram (see top right panel in Fig. \ref{fig:dexcess}), where deep, P-Cygni features are detected in a significant fraction of the spectra, while no nebular phase spectra are reported outside the $5\sigma$ confidence region. The handful of spectra classified by ATM as \textit{broad} type do lie in the nebular region of the excesses diagnostic diagram, though their inclusion within the $3\sigma$ confidence region would not trigger a detection from the traditional method alone. This might be a consequence of ATM inspecting the full line profile rather than only the wings, eventually leading to more sensitive detection. The fact that only one spectrum was classified as \textit{absorbed} suggests that the inclusion of a simulated Na\textsc{i} doublet feature in the training dataset was able to mitigate potential missclassifications to a remarkable degree.

\subsubsection{$\rm H\beta$}

Classification of these profiles reveals $48\%$ \textit{disc} spectra, $50\%$ containing some kind of outflows, and $2\%$ \textit{absorbed}. The higher ratio of \textit{disc} spectra for $\rm H\beta$, specially when compared with the almost null ratio for $\rm H\alpha$, shows that the extreme behaviour observed in the latter line is not present in $\rm H\beta$. Indeed, the Balmer decrement variability reported in MS18 is driven by the more dramatic changes in $\rm H\alpha$ (typically ${\rm EW}\sim 20- 300\,$\AA, but spiking up to ${\rm EW}\sim 2000\,$\AA), while $\rm H\beta$ is typically ${\rm EW}\sim 10-50\,$\AA \, (with certain epochs reaching up to ${\rm EW}\sim 300\,$\AA). This view is fully consistent with the reported ratio of outflow types, being $32\%$ \textit{blue-absorption}, $34\%$ \textit{broad} and $34\%$ \textit{P-Cygni}. An overall agreement can be confirmed by inspecting Fig. \ref{fig:dexcess}, bottom left panel, where the ATM classification populates the expected regions of the diagram. Finally, $2\%$ of the spectra are classified as \textit{absorbed}. Absorption features have been seen before in a number of LMXBs (see \citealt{Dubus2001} and references therein), and while they are predominantly reported in hydrogen lines, they have also been observed in He lines (e.g., \citealt{Soria2000}). A visual inspection of the affected lines in our database shows that spectra classified as \textit{absorbed} type probably arise from the presence of \textit{P-Cygni} profiles in the nearby $\rm He\textsc{i}\, 4922$ line. This might be confused with a red-shifted component for the $\rm H\beta$ line and, when combined with a blue-shifted absorption, result in the aforementioned classification. Furthermore, the detections are all restricted to the same epoch (day 6 of the outburst, see Fig. \ref{fig:realclassexample}), while typical broad absorption features are rather persistent, lasting from tens of minutes to days. For the above reasons, we believe that ATM classification for these spectra might not highlight an intrinsic feature of the system, and should be taken with caution.

\subsubsection{$\rm He\textsc{ii}\, 4686$}

ATM classification of this line profile resulted in $19\%$ \textit{disc} profiles, $77\%$ with outflow components and $4\%$ \textit{absorbed}. The large amount of detected outflow profiles is surprising for a high ionisation line. If true, it would require to explain both the absence of previous outflow reports in this line, and the fact that high densities are required for this emission line to form, which hampers the visibility of outflow features (see \citealt{Charles2019} for a discussion).

For this reason, we perform a visual inspection of the profiles and found that the detection of blue-shifted absorptions is heavily biased by the presence of the nearby Bowen blend. Indeed, this very contamination has hampered any attempts to produce an excesses diagnostic diagram for this line. This feature is a combination of various C, N, and O emission lines, with both narrow and broad components, and blended together \citep{McClintock1975}. They are thought to arise from both the companion star irradiated face through a fluorescence cascade (producing narrow emission line components), or from regions in the accretion disc where similar conditions to form the lines occur (giving rise to broader components, due to the velocity distribution; see e.g. \citealt{Steeghs2002}). For this reason, deriving a precise theoretical modelling of the Bowen blend is a challenging task. We injected into the simulated database a broad Gaussian profile to try to emulate this contaminant, but in light of the results, we do not believe it is an acceptable approximation, specially given that the Bowen blend component intensity can be comparable to that of the line of interest. For this reason, we strongly believe that any outflow detection in $\rm He\textsc{ii}\, 4686$ is probably biased, and it should be taken with the utmost caution.

\subsection{MAXI J1820}

While MAXI J1820 spectral sample is significantly smaller than that of V404 Cygni, it enables us to test ATM in a different system and using higher resolution spectra. This allows us to inspect how ATM behaves when applied to different scenarios, as well as to strengthen the conclusions derived from this novel approach when compared with traditional techniques. To aid this purpose, we show the excesses diagnostic diagram for the $\rm H\alpha$ line of the J1820 database in Fig. \ref{fig:dexcess} (bottom-right panel).

\subsubsection{$\rm H\alpha$}
The resulting classification of the $\rm H\alpha$ line shows a radically different situation to that of V404 Cygni: most of the spectra are identified as \textit{disc} ($75\%$), and the remaining ($25\%$) correspond to the \textit{broad} class. No other outflows containing blue-shifted absorptions are found, which is roughly consistent with the overall description of the data presented in MD19 and the excesses diagnostic diagram in Fig. \ref{fig:dexcess} (bottom right panel; constructed from the individual spectra instead of averaging in each epoch). A closer inspection of each particular spectrum reveals some discrepancies between ATM and the classification proposed in MD19, specially for those determined from visual inspection. 

Most of the differences occur during the initial hard state of the outburst, where outflows have been traditionally observed. First, MD19 reports on a P-Cygni profile for epoch-3 and broad wings for epoch-4; while ATM classifies all of them as \textit{disc}. We note that the original paper reports that the epoch-3 P-Cygni did not follow a traditional shape. A visual inspection confirms the profile departs from our simpler models, which would explain why ATM did not pick up the feature properly. On another vein, ATM classification of epoch-5 to -10 reveal \textit{broad} profiles, which matches the results reported in MD19 except for epoch-7 (where they report a P-Cygni component on top of the broad wings). The profile on this latter epoch is clearly dominated by its emission component, one of the strongest of all the analysed spectra. Furthermore, the confirmation of the blue-shifted absorption component in MD19 relied on the simultaneous detection of a similar outflow feature in the nearby $\rm He\textsc{i}\, 6678$ line; a piece of information ATM was not trained to identify. Together, they can explain the difference in the classification.  

\begin{figure}
\includegraphics[keepaspectratio, trim=0cm 0cm 0cm 0cm, clip=true, width=0.5\textwidth]{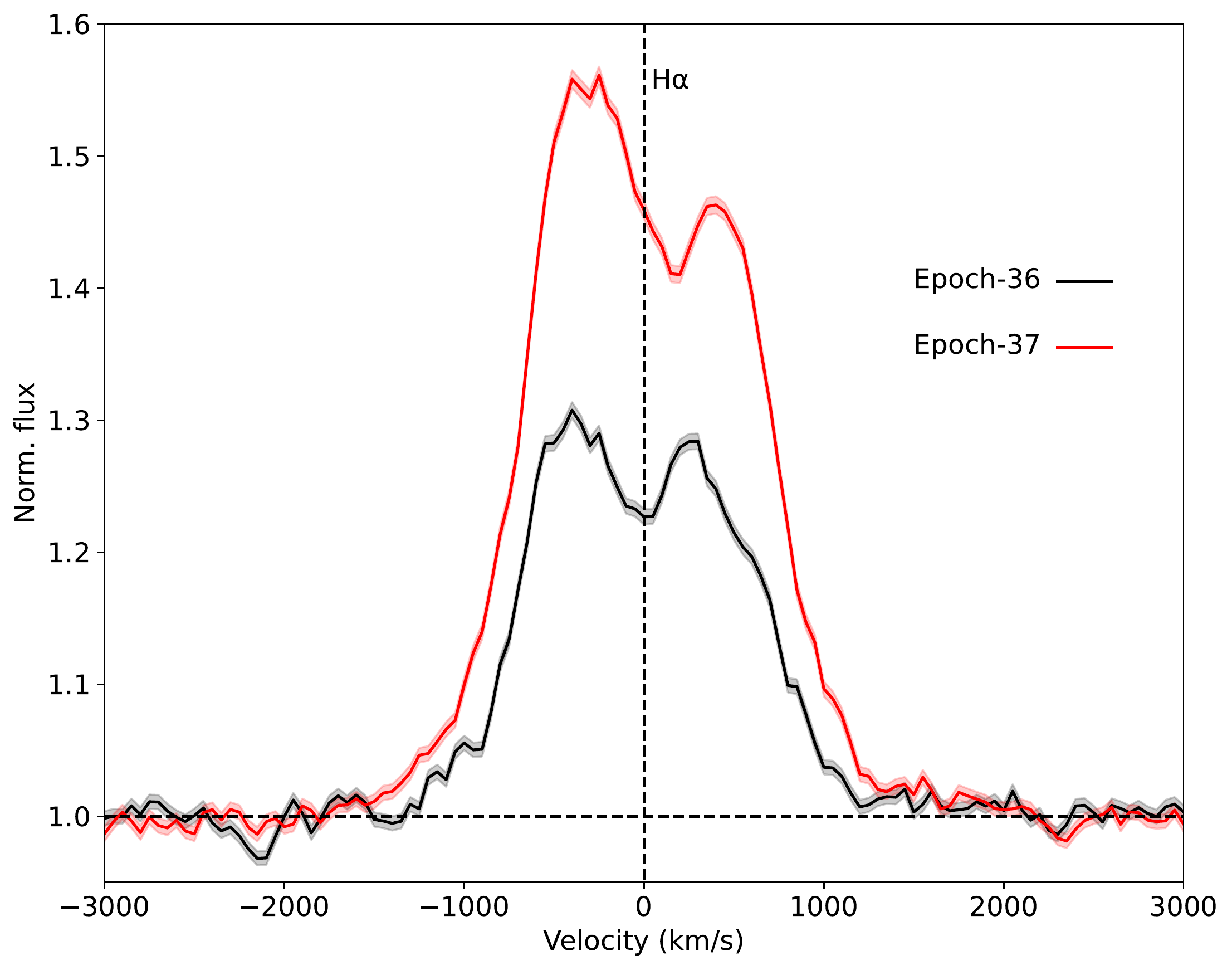}
\caption{Observed spectra of MAXI J1820 during the low luminosity hard state, corresponding to epoch-36 and -37 as defined in MD19. Colours determine the ATM classification as in previous figures, marking them as \textit{disc} and \textit{broad}, respectively.}  \label{fig:J1820spec}
\end{figure}

Epoch-11 is classified as having broad wings by MD19, but ATM suggest they can be reproduced as \textit{disc} profiles. Indeed, visual inspection of the spectra shows that the broad wings present until the previous epoch are much less prominent at epoch-11, which might either favour the results from ATM or suggest it is not able to detect such subtle features. The remaining of the dataset corresponds to soft state spectra (except for the last three low-luminosity hard state epochs), and they are classified as \textit{disc} profiles by both MD19 and ATM, providing further reassurance on the capabilities of the NN to not overestimate the presence of outflows in the data. This is consistent with the current picture of outflows not being detected during this accretion state in the optical regime (see, e.g., \citealt{SanchezSierras2020}). The only notable exception is the latest spectrum of the sample (epoch-37, hard state), where ATM reveals a \textit{broad} emission component. It is worth remarking this is the highest EW epoch of our sample, and the terminal velocities of the profile wings are similar or larger than any of the other epochs classified as possessing broad wings. Nebular phases have been observed in a handful of systems, such as the canonical V404 Cygni (MD16, MS18), GX 339-4 \citep{Rahoui2014} and more recently Aquila X-1 \citep{Panizo2021}. They are characterised by the presence of a broad emission component typically observed in $\rm H\alpha$, which becomes particularly prominent during the latest stage of the outburst decay (the so-called low hard state). Comparison with the closest spectrum in our sample (epoch-36, observed two weeks prior, but still during the hard state decay) shows that the $\rm H\alpha$ line in epoch-37 was remarkably stronger (see Fig. \ref{fig:J1820spec}). The asymmetry on the line core (redshifted from the rest wavelength) also hints towards an underlying emission component. While the feature in J1820 is not as clear as in the aforementioned works (where the single-peak nature of the line ease their detection), we favour the ATM classification in this particular case.

\subsubsection{$\rm He\textsc{i}\, 5876$}

The classification of this line profile results in $58\%$ of \textit{disc} and $40\%$ \textit{blue-absorption} (as well as a single \textit{absorbed} spectrum that we believe is due to normalisation issues). This seems consistent with the overall reported detection in the original paper, though the latter also reported two epochs of broad wing components classified as \textit{blue-absorption} profiles by ATM. Unfortunately, the excesses diagnostic diagram for this line was not reliable due to the narrowness of the line profile (with wings extending up to $<1000\, \rm{km\, s^{-1}}$), which made the red wing cross the continuum within the $\rm Na\textsc{i}$ interstellar doublet (i.e., limiting our ability to measure $\rm{EW}_{\rm r}$ values). Therefore, the original paper identification relied on visual inspection and comparison with other epochs to define them as such. However, the agnostic approach of ATM suggests that similarly broad profiles can be generated using only \textit{disc} models, as long as a hotspot contribution is included to account for the line asymmetry. We cannot confidently claim that broad wings are unambiguously detected. 

Regarding the detection of \textit{blue-absorption} profiles, they are mainly detected during the hard state, including all epochs reported in MD19. The potential detection of \textit{blue-absorption} by ATM in epoch-8 is interesting. Visual inspection of the data obtained for this epoch, however, does not show a clear blue absorption in the data. Instead, the profile for this particular date is remarkably different to those of the rest of the outburst: it consists of a single broad peak, clearly skewed towards blue wavelengths. ATM classification is actually not particularly conclusive, with a split probability of 70-30 between \textit{blue-absorption} and \textit{disc} profiles. We conclude this reflects the limitations of the simulated database to reproduce such profiles, rather than a definitive outflow feature. The remaining detections of blue absorptions by ATM correspond to ambiguous spectra obtained during the soft state. Previous detections of optical outflows have been reported during the bright soft intermediate state (see \citealt{MataSanchez2021,Panizo2022}), but they are yet to be found during the longer-lived soft state. On the other hand, the detection of NIR outflows during most of the outburst in this very system (see \citealt{SanchezSierras2020}) proves that it remains at all epochs. Indeed, visual inspection also suggests the presence of a blue absorption in our highlighted spectra. Nevertheless, the ATM probability associated with the classification is not particularly high (between $40-80\%$), which combined with the lower SNR (due to a combination of lower exposure times and brightness of the system) forbid us from reporting a definitive confirmation.

\subsection{Additional tests}

We performed a handful of tests to evaluate the performance and limitations of the ATM script defined above. A non-exhaustive list with the most remarkable among them follows:

\begin{itemize}
    \item Gaussian \textit{disc} profiles: we created an independent database where the \textit{disc} profiles are constructed through the combination of two Gaussians. The Gaussian parameters were selected such as to produce double-peaked profiles comparable to those observed in LMXBs, and the ML algorithm was trained on this new database. Regarding the performance on the separation between classes for the test dataset, the results were fundamentally similar to those obtained with the disc models described in the paper. This further strengthens the case for the re-trainability of the NN if new, more complex disc profiles are proposed in the future. Classification of the observed spectra produced different outcomes in the final classification for a number of them, while others remained the same (e.g., most of the V404 Cygni $\rm H\alpha$ spectra are still classified as being \textit{broad} or \textit{P-Cygni} types). We still favour using the disc profiles over the Gaussian approach, as the former are based on a physical description of the system, even if simplified.
    \item Universal grism approach: we explore the possibility of training the ML classifier on a single database, which can then be applied to both R1000B (V404 Cygni) and R2500R (MAXI J1820) spectra. The effect on the spectral resolution parameter is straightforward to include, as we are already accounting for a range of broadening values due to the local broadening component (see Sec. \ref{baseprofile}). The different dispersion values for each grism implies they are covering the same range of velocities with a different step size. We created a database with the thinnest step size out of the two grism setups, and allowed for a high enough range of broadening values to cover both spectral resolutions. After training the NN on this database, we applied it to the observed datasets. Application to MAXI J1820 produced fully consistent results to those presented in the previous section (using the R2500R setup), as expected due to the almost identical nature of the databases. On the other hand, application to V404 Cygni required interpolation of the observed spectrum (R1000B) to a thinner velocity grid (that of R2500R) before the ML script can be applied. The overall resulting classification is similar to that obtained with the NN trained with the adequate grism setup. However, we found a larger ratio of spectra miss-classified as \textit{absorbed} (for $\rm H\beta$ and $\rm He\textsc{i}\, 5876$). This could be an unwanted by-product of the necessary interpolation, which might create artificial shallow features out of the noise. Taking into account that this effect would be even more critical for thinner grids, we favour using a NN trained on the particular observing setup.
    \item Observational uncertainties: when dealing with observational datasets, it is reasonable to assume the inherent uncertainties will limit our outflow detection ability, as shallow features are bound to smear within the noise. This is indeed a limitation intrinsic to the observed spectra that the ML technique cannot fully overcome. We accounted for its effect by including the uncertainties into the simulated spectral flux through the SNR parameter. Nevertheless, ATM remains agnostic to the observed uncertainties when classifying a particular spectrum. We attempt to evaluate this effect by generating a set of 100 spectra using a Monte Carlo approach, employing the observed spectrum as a seed and the uncertainties on each individual flux value as the standard deviation (effectively constructing a normal distribution). The resulting spectra were then classified using the trained ML algorithm. The classification rate for each type was reported, effectively providing an uncertainty on the classification due to the observation noise. We apply this method to some of the spectra where the SNR might have contributed to a more uncertain classification, and found evenly distributed probabilities among a few classes, as expected. When applied to the remaining of the spectra, the effect was negligible and the results consistent with the original classification. Nevertheless, given that the NN was already trained with noisy simulated spectra, and that the observed spectra have the noise information indirectly engraved in their flux array, we fear double counting of the noise effect might be at play on the final classification, leading to a too conservative approach. Additionally, the NN itself provides us with a classification probability value (determined by the final softmax layer). Properly accounting for this effect requires implementation of a Bayesian approach. Nevertheless, given that the classification results were not significantly impacted (except for a handful of low SNR spectra), we decided not to pursue this approach further, and limit this work to the default analysis presented in the previous sections.

\end{itemize}

\section{Discussion}
\label{discussion}

The pilot study presented in this paper aims to explore the capabilities of ML techniques in identifying the elusive outflow features in optical spectra of LMXBs. The trained ATM algorithm produced overall consistent results with those derived via traditional methods when applied to real spectra, covering different line profiles and obtained during distinct outburst events of different LMXBs. In addition, this promising ML approach possesses a number of advantages compared with the traditional excesses diagnostic diagram technique. The latter assumes Gaussian-like wings for the accretion disc line profile. Furthermore, it is dependent on the selection of the core mask, which due to the wide variety of profiles ultimately relies on the observer's subjective decision, and it is bound to differ between systems. Finally, because of this very mask, information of the core of the line is neglected, therefore rendering it extremely difficult to detect low-velocity outflow features. On the other hand, ATM can be applied to virtually any LMXB event, given the observational setup is properly accounted for during construction of the simulated database. It inspects the full emission line profile, and it accounts for known contaminants in order to minimize false detections. Extension to any given instrument and setup configuration is straightforward, as only the spectral resolution and the dispersion of the instrument are required, which enables universal application of the script. ATM has been designed in a modular way, such that it is easy to explore different disc profiles to those presented here. That includes departures from the keplerian distribution of velocities, such as advection-dominated accretion flows (ADAFs; see e.g., \citealt{Narayan1995}), but also a more physical modelling of hotspot profiles in the disc. On the other hand, the code can be easily tweaked to include different outflow profiles (e.g. based on physical models, see \citealt{Koljonen2023}) allowing us to simulate more complex features, such as boxy profiles (e.g., \citealt{SanchezSierras2020}) or even inflows (e.g., \citealt{Cuneo2020b}). Finally, ATM can be easily extended to any line in other wavelength ranges (such as the near-infrared or the ultraviolet), as long as any contaminants are properly accounted for. Application to other accreting objects exhibiting outflows, such as cataclysmic variables or even quasi stellar objects, is also feasible, though a revision of the theoretical models would be required to accommodate each case.

At this point, it is worth highlighting a number of limitations of the ML approach that this pilot study revealed. First, a small number of spectra were miss-classified as belonging to the \textit{absorbed} class. Exploring these particular spectra revealed the origin of this effect in a combination of normalisation issues and contamination by nearby lines, which produced signatures apparently similar to broad absorptions. The most extreme example is illustrated by our attempts to classify the $\rm He\textsc{ii}\, 4686$ emission line, which, due to the presence of the nearby Bowen blend, resulted in a biased classification. Additionally, it is worth remarking that the classification of a particular spectrum is ultimately conditioned by the database employed to train the algorithm. As such, any deviations not accounted for in the profile simulation might end up generating an incorrect classification of the observed spectra. On this regard, domain adaptation techniques have proven to be able to mitigate this effect \citep{Huertas-Company2023b,Ciprijanovic2021,Ciprijanovic2023}. Finally, detection of outliers to flag observed spectra too distinct from the simulated datasets would require a complementary approach using different architectures better suited for the task (e.g. autoencoders).

The pilot study presented in this paper has proven the success and capabilities of ML techniques when applied to our particular case. Our intent is to make ATM open-source to be exploited by the community. Future plans for ATM before its public release include development of a user-friendly interface, further development to apply it to any instrumental setup configuration (e.g. creating a grid of classifiers) and the addition of a utility for outlier detection (to highlight profiles not reproduced by the simulated database).

\section{Conclusions}

We present a pilot study aiming to automatically identify outflow features in LMXB spectra using a new method based on supervised ML techniques. In order to train and test the performance of the ML algorithm, we build a simulated database containing a million spectra with line profiles based on theoretical disc models. In a fraction of these, we also injected different types of outflow features in a fraction of them. After application of three different architectures to four separate line profiles ($\rm H\alpha$, $\rm He\textsc{i}\,5876$, $\rm H\beta$ and $\rm He\textsc{ii}\, 4686$), we conclude that the best performance is attained with the ResNet classifier setup (which we name as ATM), showing an accuracy and recall over $93\%$ for all the relevant profiles. Comparison of the outflow detection and classification provided by this new technique with that attained via the traditional excesses diagnostic diagram showed overall consistent results, but with a number of associated advantages. First, we remove the dependence on the subjective determination of a mask for the analysis, as well as enable the detection of low velocity outflow features. Secondly, the underlying accretion disc model is now physically motivated, and the algorithm has been designed to ease future implementation of more complex and precise models both for the accretion disc and outflow features. We also discuss the current limitations of this technique, such as its biased performance in heavily contaminated lines or its intrinsic dependence on the theoretical models themselves. 

The notable success of ATM at classifying optical spectra for the two analysed LMXB outbursts in this pilot study (V404 Cygni and MAXI J1820) paves the way for its application to larger and more heterogeneous datasets. This tool immediately enables automatic detection and classification of outflows in future outburst events, as well as it sets the foundations for studies of the whole known population.

\section*{Acknowledgements}

DMS acknowledges support from the Consejería de Economía, Conocimiento y Empleo del Gobierno de Canarias and the European Regional Development Fund (ERDF) under grant with reference ProID2021010132 (ACCISI/FEDER, UE); as well as from the Spanish Ministry of Science and Innovation via an Europa Excelencia grant (EUR2021-122010). TMD acknowledges support via Ram\'on y Cajal Fellowship RYC-2015-18148. This work has been supported in part by the Spanish Ministry of Science under grants AYA2017-83216-P, PID2020-120323GB-I00 and EUR2021-122010. We thank Tom Marsh for the use of \textsc{molly} software. We are thankful to the GTC staff for their prompt and efficient response at triggering the time-of-opportunity program at the source of the spectroscopy presented in this work. Based on data available at the GTC Archive at CAB (CSIC -INTA). The GTC Archive is part of the Spanish Virtual Observatory project funded by MCIN/AEI/10.13039/501100011033 through grant PID2020-112949GB-I00.

\section*{Data Availability}

All data employed in this study are available at the GTC archive \url{https://gtc.sdc.cab.inta-csic.es/gtc/}. Data reduction is discussed in the original papers dedicated to the relevant sources. The simulated database employed for the training and testing of the ATM algorithm will be shared upon reasonable request to the corresponding author.
 



\bibliographystyle{mnras}
\bibliography{mybiblio} 

\begin{thebibliography}{}
\makeatletter
\relax
\def\mn@urlcharsother{\let\do\@makeother \do\$\do\&\do\#\do\^\do\_\do\%\do\~}
\def\mn@doi{\begingroup\mn@urlcharsother \@ifnextchar [ {\mn@doi@}
  {\mn@doi@[]}}
\def\mn@doi@[#1]#2{\def\@tempa{#1}\ifx\@tempa\@empty \href
  {http://dx.doi.org/#2} {doi:#2}\else \href {http://dx.doi.org/#2} {#1}\fi
  \endgroup}
\def\mn@eprint#1#2{\mn@eprint@#1:#2::\@nil}
\def\mn@eprint@arXiv#1{\href {http://arxiv.org/abs/#1} {{\tt arXiv:#1}}}
\def\mn@eprint@dblp#1{\href {http://dblp.uni-trier.de/rec/bibtex/#1.xml}
  {dblp:#1}}
\def\mn@eprint@#1:#2:#3:#4\@nil{\def\@tempa {#1}\def\@tempb {#2}\def\@tempc
  {#3}\ifx \@tempc \@empty \let \@tempc \@tempb \let \@tempb \@tempa \fi \ifx
  \@tempb \@empty \def\@tempb {arXiv}\fi \@ifundefined
  {mn@eprint@\@tempb}{\@tempb:\@tempc}{\expandafter \expandafter \csname
  mn@eprint@\@tempb\endcsname \expandafter{\@tempc}}}

\bibitem[\protect\citeauthoryear{Abadi et~al.,}{Abadi et~al.}{2015}]{Abadi2015}
Abadi M.,  et~al., 2015, {TensorFlow}: Large-Scale Machine Learning on
  Heterogeneous Systems, \url {https://www.tensorflow.org/}

\bibitem[\protect\citeauthoryear{{Atri} et~al.,}{{Atri}
  et~al.}{2019}]{Atri2019}
{Atri} P.,  et~al., 2019, \mn@doi [\mnras] {10.1093/mnras/stz2335}, \href
  {https://ui.adsabs.harvard.edu/abs/2019MNRAS.489.3116A} {489, 3116}

\bibitem[\protect\citeauthoryear{{Ball} \& {Brunner}}{{Ball} \&
  {Brunner}}{2010}]{Ball2010}
{Ball} N.~M.,  {Brunner} R.~J.,  2010, \mn@doi [International Journal of Modern
  Physics D] {10.1142/S0218271810017160}, \href
  {https://ui.adsabs.harvard.edu/abs/2010IJMPD..19.1049B} {19, 1049}

\bibitem[\protect\citeauthoryear{{Barthelmy}, {D'Ai}, {D'Avanzo}, {Krimm},
  {Lien}, {Marshall}, {Maselli}  \& {Siegel}}{{Barthelmy}
  et~al.}{2015}]{Barthelmy2015}
{Barthelmy} S.~D.,  {D'Ai} A.,  {D'Avanzo} P.,  {Krimm} H.~A.,  {Lien} A.~Y.,
  {Marshall} F.~E.,  {Maselli} A.,   {Siegel} M.~H.,  2015, GRB Coordinates
  Network, \href {https://ui.adsabs.harvard.edu/abs/2015GCN.17929....1B}
  {17929, 1}

\bibitem[\protect\citeauthoryear{{Bianchini}, {della Valle}, {Masetti}  \&
  {Margoni}}{{Bianchini} et~al.}{1997}]{Bianchini1997}
{Bianchini} A.,  {della Valle} M.,  {Masetti} N.,   {Margoni} R.,  1997, \aap,
  \href {https://ui.adsabs.harvard.edu/abs/1997A&A...321..477B} {321, 477}

\bibitem[\protect\citeauthoryear{{Brink}, {Richards}, {Poznanski}, {Bloom},
  {Rice}, {Negahban}  \& {Wainwright}}{{Brink} et~al.}{2013}]{Brink2013}
{Brink} H.,  {Richards} J.~W.,  {Poznanski} D.,  {Bloom} J.~S.,  {Rice} J.,
  {Negahban} S.,   {Wainwright} M.,  2013, \mn@doi [\mnras]
  {10.1093/mnras/stt1306}, \href
  {https://ui.adsabs.harvard.edu/abs/2013MNRAS.435.1047B} {435, 1047}

\bibitem[\protect\citeauthoryear{{Callanan} et~al.,}{{Callanan}
  et~al.}{1995}]{Callanan1995}
{Callanan} P.~J.,  et~al., 1995, \mn@doi [\apj] {10.1086/175402}, \href
  {https://ui.adsabs.harvard.edu/abs/1995ApJ...441..786C} {441, 786}

\bibitem[\protect\citeauthoryear{{Casares}, {Charles}  \& {Naylor}}{{Casares}
  et~al.}{1992}]{Casares1992}
{Casares} J.,  {Charles} P.~A.,   {Naylor} T.,  1992, \mn@doi [\nat]
  {10.1038/355614a0}, \href {http://adsabs.harvard.edu/abs/1992Natur.355..614C}
  {355, 614}

\bibitem[\protect\citeauthoryear{{Casares}, {Marsh}, {Charles}, {Martin},
  {Martin}, {Harlaftis}, {Pavlenko}  \& {Wagner}}{{Casares}
  et~al.}{1995}]{Casares1995}
{Casares} J.,  {Marsh} T.~R.,  {Charles} P.~A.,  {Martin} A.~C.,  {Martin}
  E.~L.,  {Harlaftis} E.~T.,  {Pavlenko} E.~P.,   {Wagner} R.~M.,  1995,
  \mn@doi [MNRAS] {10.1093/mnras/274.2.565}, \href
  {https://ui.adsabs.harvard.edu/abs/1995MNRAS.274..565C} {274, 565}

\bibitem[\protect\citeauthoryear{{Casares}, {Mu{\~n}oz-Darias}, {Mata
  S{\'a}nchez}, {Charles}, {Torres}, {Armas Padilla}, {Fender}  \&
  {Garc{\'\i}a-Rojas}}{{Casares} et~al.}{2019}]{Casares2019}
{Casares} J.,  {Mu{\~n}oz-Darias} T.,  {Mata S{\'a}nchez} D.,  {Charles} P.~A.,
   {Torres} M.~A.~P.,  {Armas Padilla} M.,  {Fender} R.~P.,
  {Garc{\'\i}a-Rojas} J.,  2019, \mn@doi [\mnras] {10.1093/mnras/stz1793},
  \href {https://ui.adsabs.harvard.edu/abs/2019MNRAS.488.1356C} {488, 1356}

\bibitem[\protect\citeauthoryear{{Casares} et~al.,}{{Casares}
  et~al.}{2022}]{Casares2022}
{Casares} J.,  et~al., 2022, \mn@doi [\mnras] {10.1093/mnras/stac1881}, \href
  {https://ui.adsabs.harvard.edu/abs/2022MNRAS.516.2023C} {516, 2023}

\bibitem[\protect\citeauthoryear{{Charles}, {Matthews}, {Buckley}, {Gandhi},
  {Kotze}  \& {Paice}}{{Charles} et~al.}{2019}]{Charles2019}
{Charles} P.,  {Matthews} J.~H.,  {Buckley} D. A.~H.,  {Gandhi} P.,  {Kotze}
  E.,   {Paice} J.,  2019, \mn@doi [\mnras] {10.1093/mnrasl/slz120}, \href
  {https://ui.adsabs.harvard.edu/abs/2019MNRAS.489L..47C} {489, L47}

\bibitem[\protect\citeauthoryear{Chollet et~al.}{Chollet
  et~al.}{2015}]{Chollet2015}
Chollet F.,  et~al., 2015, Keras, \url {https://github.com/fchollet/keras}

\bibitem[\protect\citeauthoryear{{{\'C}iprijanovi{\'c}}
  et~al.,}{{{\'C}iprijanovi{\'c}} et~al.}{2021}]{Ciprijanovic2021}
{{\'C}iprijanovi{\'c}} A.,  et~al., 2021, \mn@doi [\mnras]
  {10.1093/mnras/stab1677}, \href
  {https://ui.adsabs.harvard.edu/abs/2021MNRAS.506..677C} {506, 677}

\bibitem[\protect\citeauthoryear{{{\'C}iprijanovi{\'c}}, {Lewis}, {Pedro},
  {Madireddy}, {Nord}, {Perdue}  \& {Wild}}{{{\'C}iprijanovi{\'c}}
  et~al.}{2023}]{Ciprijanovic2023}
{{\'C}iprijanovi{\'c}} A.,  {Lewis} A.,  {Pedro} K.,  {Madireddy} S.,  {Nord}
  B.,  {Perdue} G.~N.,   {Wild} S.~M.,  2023, \mn@doi [Machine Learning:
  Science and Technology] {10.1088/2632-2153/acca5f}, \href
  {https://ui.adsabs.harvard.edu/abs/2023MLS&T...4b5013C} {4, 025013}

\bibitem[\protect\citeauthoryear{{Corral-Santana}, {Casares},
  {Mu{\~n}oz-Darias}, {Rodr{\'{\i}}guez-Gil}, {Shahbaz}, {Torres}, {Zurita}  \&
  {Tyndall}}{{Corral-Santana} et~al.}{2013}]{Corral-Santana2013}
{Corral-Santana} J.~M.,  {Casares} J.,  {Mu{\~n}oz-Darias} T.,
  {Rodr{\'{\i}}guez-Gil} P.,  {Shahbaz} T.,  {Torres} M.~A.~P.,  {Zurita} C.,
  {Tyndall} A.~A.,  2013, \mn@doi [Science] {10.1126/science.1228222}, \href
  {http://adsabs.harvard.edu/abs/2013Sci...339.1048C} {339, 1048}

\bibitem[\protect\citeauthoryear{{Corral-Santana}, {Casares},
  {Mu{\~n}oz-Darias}, {Bauer}, {Mart{\'{\i}}nez-Pais}  \&
  {Russell}}{{Corral-Santana} et~al.}{2016}]{Corral-Santana2016}
{Corral-Santana} J.~M.,  {Casares} J.,  {Mu{\~n}oz-Darias} T.,  {Bauer} F.~E.,
  {Mart{\'{\i}}nez-Pais} I.~G.,   {Russell} D.~M.,  2016, \mn@doi [A\&A]
  {10.1051/0004-6361/201527130}, \href
  {https://ui.adsabs.harvard.edu/abs/2016A%26A...587A..61C} {587, A61}

\bibitem[\protect\citeauthoryear{{C{\'u}neo} et~al.,}{{C{\'u}neo}
  et~al.}{2020}]{Cuneo2020b}
{C{\'u}neo} V.~A.,  et~al., 2020, \mn@doi [MNRAS] {10.1093/mnras/staa2241},
  \href {https://ui.adsabs.harvard.edu/abs/2020MNRAS.498...25C} {498, 25}

\bibitem[\protect\citeauthoryear{{D{\'\i}az Trigo} \& {Boirin}}{{D{\'\i}az
  Trigo} \& {Boirin}}{2016}]{DiazTrigo2016}
{D{\'\i}az Trigo} M.,  {Boirin} L.,  2016, \mn@doi [Astronomische Nachrichten]
  {10.1002/asna.201612315}, \href
  {https://ui.adsabs.harvard.edu/abs/2016AN....337..368D} {337, 368}

\bibitem[\protect\citeauthoryear{{Dubus}, {Kim}, {Menou}, {Szkody}  \&
  {Bowen}}{{Dubus} et~al.}{2001}]{Dubus2001}
{Dubus} G.,  {Kim} R. S.~J.,  {Menou} K.,  {Szkody} P.,   {Bowen} D.~V.,  2001,
  \mn@doi [ApJ] {10.1086/320648}, \href
  {https://ui.adsabs.harvard.edu/abs/2001ApJ...553..307D} {553, 307}

\bibitem[\protect\citeauthoryear{{Fender} \& {Mu{\~n}oz-Darias}}{{Fender} \&
  {Mu{\~n}oz-Darias}}{2016}]{Fender2016}
{Fender} R.,  {Mu{\~n}oz-Darias} T.,  2016, {The Balance of Power: Accretion
  and Feedback in Stellar Mass Black Holes}.
p.~65, \mn@doi{10.1007/978-3-319-19416-5\_3}

\bibitem[\protect\citeauthoryear{{Fender}, {Belloni}  \& {Gallo}}{{Fender}
  et~al.}{2004}]{Fender2004}
{Fender} R.~P.,  {Belloni} T.~M.,   {Gallo} E.,  2004, \mn@doi [\mnras]
  {10.1111/j.1365-2966.2004.08384.x}, \href
  {https://ui.adsabs.harvard.edu/abs/2004MNRAS.355.1105F} {355, 1105}

\bibitem[\protect\citeauthoryear{{Giacconi}, {Gursky}, {Paolini}  \&
  {Rossi}}{{Giacconi} et~al.}{1962}]{Giacconi1962}
{Giacconi} R.,  {Gursky} H.,  {Paolini} F.~R.,   {Rossi} B.~B.,  1962, \mn@doi
  [\prl] {10.1103/PhysRevLett.9.439}, \href
  {https://ui.adsabs.harvard.edu/abs/1962PhRvL...9..439G} {9, 439}

\bibitem[\protect\citeauthoryear{{Horne} \& {Marsh}}{{Horne} \&
  {Marsh}}{1986}]{Horne1986}
{Horne} K.,  {Marsh} T.~R.,  1986, \mn@doi [MNRAS] {10.1093/mnras/218.4.761},
  \href {https://ui.adsabs.harvard.edu/abs/1986MNRAS.218..761H} {218, 761}

\bibitem[\protect\citeauthoryear{{Huertas-Company} \&
  {Lanusse}}{{Huertas-Company} \& {Lanusse}}{2023}]{Huertas-Company2023}
{Huertas-Company} M.,  {Lanusse} F.,  2023, \mn@doi [\pasa]
  {10.1017/pasa.2022.55}, \href
  {https://ui.adsabs.harvard.edu/abs/2023PASA...40....1H} {40, e001}

\bibitem[\protect\citeauthoryear{{Huertas-Company} et~al.,}{{Huertas-Company}
  et~al.}{2023}]{Huertas-Company2023b}
{Huertas-Company} M.,  et~al., 2023, \mn@doi [arXiv e-prints]
  {10.48550/arXiv.2305.02478}, \href
  {https://ui.adsabs.harvard.edu/abs/2023arXiv230502478H} {p. arXiv:2305.02478}

\bibitem[\protect\citeauthoryear{{Iijima} \& {Esenoglu}}{{Iijima} \&
  {Esenoglu}}{2003}]{Iijima2003}
{Iijima} T.,  {Esenoglu} H.~H.,  2003, \mn@doi [\aap]
  {10.1051/0004-6361:20030528}, \href
  {https://ui.adsabs.harvard.edu/abs/2003A&A...404..997I} {404, 997}

\bibitem[\protect\citeauthoryear{Ioffe \& Szegedy}{Ioffe \&
  Szegedy}{2015}]{Ioffe2015}
Ioffe S.,  Szegedy C.,  2015, in Bach F.,  Blei D.,  eds,  Proceedings of
  Machine Learning Research Vol. 37, Proceedings of the 32nd International
  Conference on Machine Learning. PMLR, Lille, France, pp 448--456, \url
  {https://proceedings.mlr.press/v37/ioffe15.html}

\bibitem[\protect\citeauthoryear{{Ismail Fawaz}, {Forestier}, {Weber},
  {Idoumghar}  \& {Muller}}{{Ismail Fawaz} et~al.}{2018}]{Fawaz2018}
{Ismail Fawaz} H.,  {Forestier} G.,  {Weber} J.,  {Idoumghar} L.,   {Muller}
  P.-A.,  2018, \mn@doi [Data Mining and Knowledge Discovery]
  {10.1007/s10618-019-00619-1}, \href
  {https://ui.adsabs.harvard.edu/abs/2018arXiv180904356I} {33, 3116}

\bibitem[\protect\citeauthoryear{{Jim{\'e}nez-Ibarra}, {Mu{\~n}oz-Darias},
  {Casares}, {Armas Padilla}  \& {Corral-Santana}}{{Jim{\'e}nez-Ibarra}
  et~al.}{2019}]{JimenezIbarra2019b}
{Jim{\'e}nez-Ibarra} F.,  {Mu{\~n}oz-Darias} T.,  {Casares} J.,  {Armas
  Padilla} M.,   {Corral-Santana} J.~M.,  2019, \mn@doi [MNRAS]
  {10.1093/mnras/stz2393}, \href
  {https://ui.adsabs.harvard.edu/abs/2019MNRAS.489.3420J} {489, 3420}

\bibitem[\protect\citeauthoryear{{Killestein} et~al.,}{{Killestein}
  et~al.}{2021}]{Killestein2021}
{Killestein} T.~L.,  et~al., 2021, \mn@doi [\mnras] {10.1093/mnras/stab633},
  \href {https://ui.adsabs.harvard.edu/abs/2021MNRAS.503.4838K} {503, 4838}

\bibitem[\protect\citeauthoryear{{Kimura} et~al.,}{{Kimura}
  et~al.}{2016}]{Kimura2016}
{Kimura} M.,  et~al., 2016, \mn@doi [\nat] {10.1038/nature16452}, \href
  {https://ui.adsabs.harvard.edu/abs/2016Natur.529...54K} {529, 54}

\bibitem[\protect\citeauthoryear{{King}, {Miller}, {Raymond}, {Reynolds}  \&
  {Morningstar}}{{King} et~al.}{2015}]{King2015}
{King} A.~L.,  {Miller} J.~M.,  {Raymond} J.,  {Reynolds} M.~T.,
  {Morningstar} W.,  2015, \mn@doi [\apjl] {10.1088/2041-8205/813/2/L37}, \href
  {https://ui.adsabs.harvard.edu/abs/2015ApJ...813L..37K} {813, L37}

\bibitem[\protect\citeauthoryear{{Koljonen}, {Long}, {Matthews}  \&
  {Knigge}}{{Koljonen} et~al.}{2023}]{Koljonen2023}
{Koljonen} K.~I.~I.,  {Long} K.~S.,  {Matthews} J.~H.,   {Knigge} C.,  2023,
  \mn@doi [\mnras] {10.1093/mnras/stad809}, \href
  {https://ui.adsabs.harvard.edu/abs/2023MNRAS.tmp..796K} {}

\bibitem[\protect\citeauthoryear{Krizhevsky, Sutskever  \& Hinton}{Krizhevsky
  et~al.}{2017}]{Krizhevsky2012}
Krizhevsky A.,  Sutskever I.,   Hinton G.~E.,  2017, \mn@doi [Commun. ACM]
  {10.1145/3065386}, 60, 84–90

\bibitem[\protect\citeauthoryear{{Liu}, {van Paradijs}  \& {van den
  Heuvel}}{{Liu} et~al.}{2007}]{Liu2007}
{Liu} Q.~Z.,  {van Paradijs} J.,   {van den Heuvel} E.~P.~J.,  2007, \mn@doi
  [A\&A] {10.1051/0004-6361:20077303}, \href
  {https://ui.adsabs.harvard.edu/abs/2007A&A...469..807L} {469, 807}

\bibitem[\protect\citeauthoryear{{Masetti}, {Bianchini}  \& {della
  Valle}}{{Masetti} et~al.}{1997}]{Masetti1997}
{Masetti} N.,  {Bianchini} A.,   {della Valle} M.,  1997, \aap, \href
  {https://ui.adsabs.harvard.edu/abs/1997A&A...317..769M} {317, 769}

\bibitem[\protect\citeauthoryear{{Mata S{\'a}nchez}, {Mu{\~n}oz-Darias},
  {Casares}, {Corral-Santana}  \& {Shahbaz}}{{Mata S{\'a}nchez}
  et~al.}{2015}]{MataSanchez2015b}
{Mata S{\'a}nchez} D.,  {Mu{\~n}oz-Darias} T.,  {Casares} J.,  {Corral-Santana}
  J.~M.,   {Shahbaz} T.,  2015, \mn@doi [MNRAS] {10.1093/mnras/stv2111}, \href
  {http://adsabs.harvard.edu/abs/2015MNRAS.454.2199M} {454, 2199}

\bibitem[\protect\citeauthoryear{{Mata S{\'a}nchez} et~al.,}{{Mata S{\'a}nchez}
  et~al.}{2018}]{MataSanchez2018}
{Mata S{\'a}nchez} D.,  et~al., 2018, \mn@doi [MNRAS] {10.1093/mnras/sty2402},
  \href {https://ui.adsabs.harvard.edu/abs/2018MNRAS.481.2646M} {481, 2646}

\bibitem[\protect\citeauthoryear{{Mata S{\'a}nchez}, {Rau}, {{\'A}lvarez
  Hern{\'a}ndez}, {van Grunsven}, {Torres}  \& {Jonker}}{{Mata S{\'a}nchez}
  et~al.}{2021}]{MataSanchez2021}
{Mata S{\'a}nchez} D.,  {Rau} A.,  {{\'A}lvarez Hern{\'a}ndez} A.,  {van
  Grunsven} T.~F.~J.,  {Torres} M.~A.~P.,   {Jonker} P.~G.,  2021, \mn@doi
  [MNRAS] {10.1093/mnras/stab1714}, \href
  {https://ui.adsabs.harvard.edu/abs/2021MNRAS.506..581M} {506, 581}

\bibitem[\protect\citeauthoryear{{Mata S{\'a}nchez} et~al.,}{{Mata S{\'a}nchez}
  et~al.}{2022}]{MataSanchez2022}
{Mata S{\'a}nchez} D.,  et~al., 2022, \mn@doi [\apjl]
  {10.3847/2041-8213/ac502f}, \href
  {https://ui.adsabs.harvard.edu/abs/2022ApJ...926L..10M} {926, L10}

\bibitem[\protect\citeauthoryear{{McClintock}, {Canizares}  \&
  {Tarter}}{{McClintock} et~al.}{1975}]{McClintock1975}
{McClintock} J.~E.,  {Canizares} C.~R.,   {Tarter} C.~B.,  1975, \mn@doi [\apj]
  {10.1086/153642}, \href
  {https://ui.adsabs.harvard.edu/abs/1975ApJ...198..641M} {198, 641}

\bibitem[\protect\citeauthoryear{{Mikolov}, {Sutskever}, {Chen}, {Corrado}  \&
  {Dean}}{{Mikolov} et~al.}{2013}]{Mikolov2013}
{Mikolov} T.,  {Sutskever} I.,  {Chen} K.,  {Corrado} G.,   {Dean} J.,  2013,
  \mn@doi [arXiv e-prints] {10.48550/arXiv.1310.4546}, \href
  {https://ui.adsabs.harvard.edu/abs/2013arXiv1310.4546M} {p. arXiv:1310.4546}

\bibitem[\protect\citeauthoryear{{Motta}, {Kajava},
  {S{\'a}nchez-Fern{\'a}ndez}, {Giustini}  \& {Kuulkers}}{{Motta}
  et~al.}{2017}]{Motta2017}
{Motta} S.~E.,  {Kajava} J.~J.~E.,  {S{\'a}nchez-Fern{\'a}ndez} C.,  {Giustini}
  M.,   {Kuulkers} E.,  2017, \mn@doi [\mnras] {10.1093/mnras/stx466}, \href
  {https://ui.adsabs.harvard.edu/abs/2017MNRAS.468..981M} {468, 981}

\bibitem[\protect\citeauthoryear{{Mu{\~n}oz-Darias} \&
  {Ponti}}{{Mu{\~n}oz-Darias} \& {Ponti}}{2022}]{Munoz-Darias2022}
{Mu{\~n}oz-Darias} T.,  {Ponti} G.,  2022, arXiv e-prints, \href
  {https://ui.adsabs.harvard.edu/abs/2022arXiv220514162M} {p. arXiv:2205.14162}

\bibitem[\protect\citeauthoryear{{Mu{\~n}oz-Darias} et~al.,}{{Mu{\~n}oz-Darias}
  et~al.}{2016}]{Munoz-Darias2016}
{Mu{\~n}oz-Darias} T.,  et~al., 2016, \mn@doi [\nat] {10.1038/nature17446},
  \href {https://ui.adsabs.harvard.edu/abs/2016Natur.534...75M} {534, 75}

\bibitem[\protect\citeauthoryear{{Mu{\~n}oz-Darias} et~al.,}{{Mu{\~n}oz-Darias}
  et~al.}{2019}]{Munoz-Darias2019}
{Mu{\~n}oz-Darias} T.,  et~al., 2019, \mn@doi [ApJ] {10.3847/2041-8213/ab2768},
  \href {https://ui.adsabs.harvard.edu/abs/2019ApJ...879L...4M} {879, L4}

\bibitem[\protect\citeauthoryear{{Narayan} \& {Yi}}{{Narayan} \&
  {Yi}}{1995}]{Narayan1995}
{Narayan} R.,  {Yi} I.,  1995, \mn@doi [\apj] {10.1086/176343}, \href
  {https://ui.adsabs.harvard.edu/abs/1995ApJ...452..710N} {452, 710}

\bibitem[\protect\citeauthoryear{{Neilsen} \& {Lee}}{{Neilsen} \&
  {Lee}}{2009}]{Neilsen2009}
{Neilsen} J.,  {Lee} J.~C.,  2009, \mn@doi [\nat] {10.1038/nature07680}, \href
  {https://ui.adsabs.harvard.edu/abs/2009Natur.458..481N} {458, 481}

\bibitem[\protect\citeauthoryear{{Orosz} \& {Bailyn}}{{Orosz} \&
  {Bailyn}}{1995}]{Orosz1995}
{Orosz} J.~A.,  {Bailyn} C.~D.,  1995, \mn@doi [ApJ] {10.1086/187930}, \href
  {http://adsabs.harvard.edu/abs/1995ApJ...446L..59O} {446, L59}

\bibitem[\protect\citeauthoryear{{Orosz}, {Bailyn}, {Remillard}, {McClintock}
  \& {Foltz}}{{Orosz} et~al.}{1994}]{Orosz1994}
{Orosz} J.~A.,  {Bailyn} C.~D.,  {Remillard} R.~A.,  {McClintock} J.~E.,
  {Foltz} C.~B.,  1994, \mn@doi [ApJ] {10.1086/174962}, \href
  {http://adsabs.harvard.edu/abs/1994ApJ...436..848O} {436, 848}

\bibitem[\protect\citeauthoryear{{Orwat-Kapola}, {Bird}, {Hill}, {Altamirano}
  \& {Huppenkothen}}{{Orwat-Kapola} et~al.}{2022}]{OrwatKapola2022}
{Orwat-Kapola} J.~K.,  {Bird} A.~J.,  {Hill} A.~B.,  {Altamirano} D.,
  {Huppenkothen} D.,  2022, \mn@doi [\mnras] {10.1093/mnras/stab3043}, \href
  {https://ui.adsabs.harvard.edu/abs/2022MNRAS.509.1269O} {509, 1269}

\bibitem[\protect\citeauthoryear{{Panizo-Espinar}, {Mu{\~n}oz-Darias}, {Armas
  Padilla}, {Jim{\'e}nez-Ibarra}, {Casares}  \& {Mata
  S{\'a}nchez}}{{Panizo-Espinar} et~al.}{2021}]{Panizo2021}
{Panizo-Espinar} G.,  {Mu{\~n}oz-Darias} T.,  {Armas Padilla} M.,
  {Jim{\'e}nez-Ibarra} F.,  {Casares} J.,   {Mata S{\'a}nchez} D.,  2021,
  \mn@doi [A\&A] {10.1051/0004-6361/202140323}, \href
  {https://ui.adsabs.harvard.edu/abs/2021A&A...650A.135P} {650, A135}

\bibitem[\protect\citeauthoryear{{Panizo-Espinar} et~al.,}{{Panizo-Espinar}
  et~al.}{2022}]{Panizo2022}
{Panizo-Espinar} G.,  et~al., 2022, \mn@doi [\aap]
  {10.1051/0004-6361/202243426}, \href
  {https://ui.adsabs.harvard.edu/abs/2022A&A...664A.100P} {664, A100}

\bibitem[\protect\citeauthoryear{{Pattnaik}, {Sharma}, {Alabarta},
  {Altamirano}, {Chakraborty}, {Kembhavi}, {M{\'e}ndez}  \&
  {Orwat-Kapola}}{{Pattnaik} et~al.}{2021}]{Pattnaik2021}
{Pattnaik} R.,  {Sharma} K.,  {Alabarta} K.,  {Altamirano} D.,  {Chakraborty}
  M.,  {Kembhavi} A.,  {M{\'e}ndez} M.,   {Orwat-Kapola} J.~K.,  2021, \mn@doi
  [\mnras] {10.1093/mnras/staa3899}, \href
  {https://ui.adsabs.harvard.edu/abs/2021MNRAS.501.3457P} {501, 3457}

\bibitem[\protect\citeauthoryear{{Ponti}, {Fender}, {Begelman}, {Dunn},
  {Neilsen}  \& {Coriat}}{{Ponti} et~al.}{2012}]{Ponti2012}
{Ponti} G.,  {Fender} R.~P.,  {Begelman} M.~C.,  {Dunn} R.~J.~H.,  {Neilsen}
  J.,   {Coriat} M.,  2012, \mn@doi [MNRAS] {10.1111/j.1745-3933.2012.01224.x},
  \href {https://ui.adsabs.harvard.edu/abs/2012MNRAS.422L..11P} {422, L11}

\bibitem[\protect\citeauthoryear{{Ponti}, {Bianchi}, {Mu{\~n}oz-Darias}, {De},
  {Fender}  \& {Merloni}}{{Ponti} et~al.}{2016}]{Ponti2016}
{Ponti} G.,  {Bianchi} S.,  {Mu{\~n}oz-Darias} T.,  {De} K.,  {Fender} R.,
  {Merloni} A.,  2016, \mn@doi [Astronomische Nachrichten]
  {10.1002/asna.201612339}, \href
  {https://ui.adsabs.harvard.edu/abs/2016AN....337..512P} {337, 512}

\bibitem[\protect\citeauthoryear{{Rahoui}, {Coriat}  \& {Lee}}{{Rahoui}
  et~al.}{2014}]{Rahoui2014}
{Rahoui} F.,  {Coriat} M.,   {Lee} J.~C.,  2014, \mn@doi [MNRAS]
  {10.1093/mnras/stu977}, \href
  {https://ui.adsabs.harvard.edu/abs/2014MNRAS.442.1610R} {442, 1610}

\bibitem[\protect\citeauthoryear{{Ricketts}, {Steiner}, {Garraffo}, {Remillard}
   \& {Huppenkothen}}{{Ricketts} et~al.}{2023}]{Ricketts2023}
{Ricketts} B.~J.,  {Steiner} J.~F.,  {Garraffo} C.,  {Remillard} R.~A.,
  {Huppenkothen} D.,  2023, arXiv e-prints, \href
  {https://ui.adsabs.harvard.edu/abs/2023arXiv230110467R} {p. arXiv:2301.10467}

\bibitem[\protect\citeauthoryear{Sainath, Mohamed, Kingsbury  \&
  Ramabhadran}{Sainath et~al.}{2013}]{Sainath2013}
Sainath T.~N.,  Mohamed A.-r.,  Kingsbury B.,   Ramabhadran B.,  2013, in 2013
  IEEE International Conference on Acoustics, Speech and Signal Processing. pp
  8614--8618, \mn@doi{10.1109/ICASSP.2013.6639347}

\bibitem[\protect\citeauthoryear{{S{\'a}nchez-Sierras} \&
  {Mu{\~n}oz-Darias}}{{S{\'a}nchez-Sierras} \&
  {Mu{\~n}oz-Darias}}{2020}]{SanchezSierras2020}
{S{\'a}nchez-Sierras} J.,  {Mu{\~n}oz-Darias} T.,  2020, \mn@doi [A\&A]
  {10.1051/0004-6361/202038406}, \href
  {https://ui.adsabs.harvard.edu/abs/2020A&A...640L...3S} {640, L3}

\bibitem[\protect\citeauthoryear{{Shrader}, {Wagner}, {Charles}, {Harlaftis}
  \& {Naylor}}{{Shrader} et~al.}{1997}]{Shrader1997}
{Shrader} C.~R.,  {Wagner} R.~M.,  {Charles} P.~A.,  {Harlaftis} E.~T.,
  {Naylor} T.,  1997, \mn@doi [\apj] {10.1086/304635}, \href
  {https://ui.adsabs.harvard.edu/abs/1997ApJ...487..858S} {487, 858}

\bibitem[\protect\citeauthoryear{Smith \& Hartigan}{Smith \&
  Hartigan}{2006}]{Smith2006}
Smith N.,  Hartigan P.,  2006, \mn@doi [The Astrophysical Journal]
  {10.1086/498860}, 638, 1045

\bibitem[\protect\citeauthoryear{{Smith}, {Mauerhan}  \& {Prieto}}{{Smith}
  et~al.}{2014}]{Smith2014}
{Smith} N.,  {Mauerhan} J.~C.,   {Prieto} J.~L.,  2014, \mn@doi [\mnras]
  {10.1093/mnras/stt2269}, \href
  {https://ui.adsabs.harvard.edu/abs/2014MNRAS.438.1191S} {438, 1191}

\bibitem[\protect\citeauthoryear{{Soria}, {Wu}  \& {Hunstead}}{{Soria}
  et~al.}{2000}]{Soria2000}
{Soria} R.,  {Wu} K.,   {Hunstead} R.~W.,  2000, \mn@doi [\apj]
  {10.1086/309194}, \href
  {https://ui.adsabs.harvard.edu/abs/2000ApJ...539..445S} {539, 445}

\bibitem[\protect\citeauthoryear{Srivastava, Hinton, Krizhevsky, Sutskever  \&
  Salakhutdinov}{Srivastava et~al.}{2014}]{Srivastava2014}
Srivastava N.,  Hinton G.,  Krizhevsky A.,  Sutskever I.,   Salakhutdinov R.,
  2014, Journal of Machine Learning Research, 15, 1929

\bibitem[\protect\citeauthoryear{{Steeghs} \& {Casares}}{{Steeghs} \&
  {Casares}}{2002}]{Steeghs2002}
{Steeghs} D.,  {Casares} J.,  2002, \mn@doi [ApJ] {10.1086/339224}, \href
  {http://adsabs.harvard.edu/abs/2002ApJ...568..273S} {568, 273}

\bibitem[\protect\citeauthoryear{Szegedy et~al.,}{Szegedy
  et~al.}{2015}]{Szegedy2015}
Szegedy C.,  et~al., 2015, in 2015 IEEE Conference on Computer Vision and
  Pattern Recognition (CVPR). pp~1--9, \mn@doi{10.1109/CVPR.2015.7298594}

\bibitem[\protect\citeauthoryear{{Tetarenko}, {Sivakoff}, {Heinke}  \&
  {Gladstone}}{{Tetarenko} et~al.}{2016}]{Tetarenko2016}
{Tetarenko} B.~E.,  {Sivakoff} G.~R.,  {Heinke} C.~O.,   {Gladstone} J.~C.,
  2016, \mn@doi [\apjs] {10.3847/0067-0049/222/2/15}, \href
  {https://ui.adsabs.harvard.edu/abs/2016ApJS..222...15T} {222, 15}

\bibitem[\protect\citeauthoryear{{Th{\"o}ne} et~al.,}{{Th{\"o}ne}
  et~al.}{2017}]{Thone2017}
{Th{\"o}ne} C.~C.,  et~al., 2017, \mn@doi [\aap] {10.1051/0004-6361/201629968},
  \href {https://ui.adsabs.harvard.edu/abs/2017A&A...599A.129T} {599, A129}

\bibitem[\protect\citeauthoryear{{Torres}, {Jonker}, {Miller-Jones}, {Steeghs},
  {Repetto}  \& {Wu}}{{Torres} et~al.}{2015}]{Torres2015}
{Torres} M.~A.~P.,  {Jonker} P.~G.,  {Miller-Jones} J.~C.~A.,  {Steeghs} D.,
  {Repetto} S.,   {Wu} J.,  2015, \mn@doi [MNRAS] {10.1093/mnras/stv720}, \href
  {http://adsabs.harvard.edu/abs/2015MNRAS.450.4292T} {450, 4292}

\bibitem[\protect\citeauthoryear{{Torres}, {Casares}, {Jim{\'e}nez-Ibarra},
  {Mu{\~n}oz-Darias}, {Armas Padilla}, {Jonker}  \& {Heida}}{{Torres}
  et~al.}{2019}]{Torres2019}
{Torres} M.~A.~P.,  {Casares} J.,  {Jim{\'e}nez-Ibarra} F.,  {Mu{\~n}oz-Darias}
  T.,  {Armas Padilla} M.,  {Jonker} P.~G.,   {Heida} M.,  2019, \mn@doi [ApJ]
  {10.3847/2041-8213/ab39df}, \href
  {https://ui.adsabs.harvard.edu/abs/2019ApJ...882L..21T} {882, L21}

\bibitem[\protect\citeauthoryear{{Torres}, {Casares}, {Jim{\'e}nez-Ibarra},
  {{\'A}lvarez-Hern{\'a}ndez}, {Mu{\~n}oz-Darias}, {Armas Padilla}, {Jonker}
  \& {Heida}}{{Torres} et~al.}{2020}]{Torres2020}
{Torres} M.~A.~P.,  {Casares} J.,  {Jim{\'e}nez-Ibarra} F.,
  {{\'A}lvarez-Hern{\'a}ndez} A.,  {Mu{\~n}oz-Darias} T.,  {Armas Padilla} M.,
  {Jonker} P.~G.,   {Heida} M.,  2020, \mn@doi [ApJ]
  {10.3847/2041-8213/ab863a}, \href
  {https://ui.adsabs.harvard.edu/abs/2020ApJ...893L..37T} {893, L37}

\bibitem[\protect\citeauthoryear{{Tucker} et~al.,}{{Tucker}
  et~al.}{2018}]{Tucker2018}
{Tucker} M.~A.,  et~al., 2018, \mn@doi [\apjl] {10.3847/2041-8213/aae88a},
  \href {https://ui.adsabs.harvard.edu/abs/2018ApJ...867L...9T} {867, L9}

\bibitem[\protect\citeauthoryear{Wang, Yan  \& Oates}{Wang
  et~al.}{2017}]{Wang2017b}
Wang Z.,  Yan W.,   Oates T.,  2017, in 2017 International Joint Conference on
  Neural Networks (IJCNN). pp 1578--1585, \mn@doi{10.1109/IJCNN.2017.7966039}

\bibitem[\protect\citeauthoryear{{Wright} et~al.,}{{Wright}
  et~al.}{2015}]{Wright2015}
{Wright} D.~E.,  et~al., 2015, \mn@doi [\mnras] {10.1093/mnras/stv292}, \href
  {https://ui.adsabs.harvard.edu/abs/2015MNRAS.449..451W} {449, 451}

\bibitem[\protect\citeauthoryear{{de Beurs}, {Islam}, {Gopalan}  \&
  {Vrtilek}}{{de Beurs} et~al.}{2022}]{deBeurs2022}
{de Beurs} Z.~L.,  {Islam} N.,  {Gopalan} G.,   {Vrtilek} S.~D.,  2022, \mn@doi
  [\apj] {10.3847/1538-4357/ac6184}, \href
  {https://ui.adsabs.harvard.edu/abs/2022ApJ...933..116D} {933, 116}

\bibitem[\protect\citeauthoryear{{della Valle}, {Jarvis}  \& {West}}{{della
  Valle} et~al.}{1991}]{dellaValle1991}
{della Valle} M.,  {Jarvis} B.~J.,   {West} R.~M.,  1991, \mn@doi [\nat]
  {10.1038/353050a0}, \href
  {https://ui.adsabs.harvard.edu/abs/1991Natur.353...50D} {353, 50}

\bibitem[\protect\citeauthoryear{{della Valle}, {Benetti}, {Cappellaro}  \&
  {Wheeler}}{{della Valle} et~al.}{1997}]{dellavalle1997}
{della Valle} M.,  {Benetti} S.,  {Cappellaro} E.,   {Wheeler} C.,  1997, \aap,
  \href {https://ui.adsabs.harvard.edu/abs/1997A&A...318..179D} {318, 179}

\makeatother
\end{thebibliography}








\bsp	
\label{lastpage}
\end{document}